\documentclass[12pt]{article}

\def\eptwo{\left\{ \phantom{|}^{\mu\nu}_{ab} \right\}}
\def\epthree{\left\{ \phantom{|}^{\mu\nu\alpha}_{abc} \right\}}

\title{On gravitational interactions \\
for massive higher spins in $AdS_3$}

\author{I.L. Buchbinder${}^{a}$\thanks{joseph@tspu.edu.ru},
T.V. Snegirev${}^a$\thanks{snegirev@tspu.edu.ru}, Yu.M.
Zinoviev$^b$\thanks{Yurii.Zinoviev@ihep.ru}
\\[0.5cm]
\it ${}^a$Department of Theoretical Physics,\\
\it Tomsk State Pedagogical University,\\
\it Tomsk 634061, Russia\\[0.3cm]
\it ${}^b$Institute for High Energy Physics,\\
\it Protvino, Moscow Region, 142280, Russia}

\date{}

\begin{document}

\maketitle

\begin{abstract}
In this paper we investigate gravitational interactions of massive
higher spin fields in three dimensional $AdS$ space with arbitrary
value of cosmological constant including flat Minkowski space. We
use frame-like gauge description for such massive fields adopted to
three-dimensional case. At first, we carefully analyze the procedure
of switching on gravitational interactions in the linear
approximation on the example of massive spin-3 field and then
proceed with the generalization to the case of arbitrary integer
spin field. As a result we construct a cubic interaction vertex
linear in spin-2 field and quadratic in higher spin field on $AdS_3$
background. As in the massless case the vertex does not contain any
higher derivative corrections to the Lagrangian and/or gauge
transformations. Thus, even after switching on gravitational
interactions, one can freely consider any massless or partially
massless limits as well as the flat one.

\end{abstract}

\thispagestyle{empty}
\newpage
\setcounter{page}{1}

\section*{Introduction}

Three dimensional higher spin field theories that appear to be much
simpler than the higher dimensional ones (for some reviews on higher
spin theories see e.g. \cite{Vas04,Sor04,BCIV05,FT08,BBS10}) may
provide nice playground to gain some useful experience. One of the
main reasons is that, contrary to the situation in $d \ge 4$
dimensions \cite{Vas90,Vas96,Vas03}, in three dimensions it is not
necessary to consider infinite number of fields to construct
consistent closed theory \cite{AD84,Ble89,CFPT10}. Also it is
important that in the frame-like formalism many such theories can be
considered as Chern-Simons ones corresponding to some non-compact
algebras \cite{Wit88,Ble89,CFPT10},

Till now most of works on the three dimensional higher spin filed
theories were devoted to construction of interaction for massless
fields \cite{PV98,PSV99,CFPT10} or parity odd topologically massive
ones \cite{CLW11,BLSS11,CL11}. But massless higher spin fields in
three dimensions being pure gauge do not have any physical degrees
of freedom, while topologically massive one though do contain
physical degrees of freedom but due to their specific properties can
hardly be generalized to higher dimensions. That is why we think
that it is important to investigate parity even interacting theories
for massive higher spin fields\footnote{Lagrangian formulation for
$d=3$ massive higher spin fields has been given for the first time in
\cite{TV}. The aspects of massive higher spin fields in $AdS_3$ are
also discussed in \cite{M} and \cite{F}.}. On the one hand, such
theories will certainly have physical interest by themselves, e.g.
for study the various aspects of duality, and from the other hand we
may expect that they admit higher dimensional generalizations. And
it seems natural to begin such investigations with the gravitational
interactions that play fundamental role in any higher spin theory.

In this paper we consider gravitational interactions for massive
higher spin fields in $AdS_3$ space in the linear approximation. It
means that we construct a cubic interacting vertex quadratic in
massive higher spin field and linear in gravitational field on
$AdS_3$ background. We use frame-like gauge invariant formalism for
massive fields \cite{Zin08b,PV10} adopted to three dimensions
elaborated in our recent paper \cite{BSZ12a}. To our opinion such
formalism is the most convenient one for investigations of massive
higher spin interactions. From one hand, its gauge invariance allows
one to extend constructive approach to investigation of massless
higher spin interactions to the case of arbitrary combination of
massless and/or massive one with the interaction of massless gravity
with massive higher spins being one important example. From the
other hand, such formalism nicely works in $AdS$ space with
arbitrary value of cosmological constant including flat Minkowski
space and this allows one to consider all possible massless and
partially massless limits that exist for given field
\cite{DW01a,DW01c,Zin01,Zin08b} as well as the flat limit.

The work is organized as follows. In Section 1 as an illustration of
general technics that we will use for massive fields in subsequent
sections we provide a couple of simplest examples of interacting
massless theories. Namely, we consider massless $d=3$ gravity itself
and the model of its interaction with massless spin-3 field for the
first time constructed in \cite{CFPT10}. Then in Section 2 we
carefully analyze gravitational interactions for massive spin 3
field. The main result here that even in massive case it is possible
to switch on gravitational interaction in the linear approximation,
e.g. to construct the cubic vertex linear in gravitational field and
quadratic in massive higher spin field, without any need to
introduce higher derivative corrections to the Lagrangian and/or
gauge transformations typical for higher dimensional theories
\cite{FV87,FV87a}. This, in turn, implies that even after switching on
gravitational interactions (at least in the linear approximation) one
can freely consider any massless or partially massless limits as well
as the flat one. At the same time we argue that contrary to the
massless case the system gravity plus massive spin-3 is not closed
so that to construct closed consistent theory one has to introduce
other fields and/or interactions. Than in Section 3 we generalize
these results to the case of arbitrary integer spin and obtain the
corresponding cubic vertex. As in the spin-3 case, this vertex also
does not contain any higher derivative corrections and admit
non-singular massless, partially massless and flat limits. It was
clear from the very beginning that the system under consideration can
not be
closed (simply because massless theory is not closed) but let us
stress that that the results of linear approximation do not depend on
the presence of any other fields so that the structure of cubic vertex
is model independent.

{\bf Notations and conventions.} We use Greek letters for the world
indices and Latin letters for the local ones. We work in $(A)dS$
space with arbitrary value of cosmological constant $\Lambda$ and use
the notation $D_\mu$ for $AdS$ covariant derivative normalized so that
$$
[D_\mu,D_\nu] \xi^a = \lambda^2 e_{[\mu}{}^a \xi_{\nu]},
\qquad \lambda^2 = - \Lambda
$$
where $e_\mu{}^a$ plays the role of (non-dynamical) frame of $(A)dS$
background. Also to write the expression in totally antisymmetric form
on the world indices (which is equivalent to an external product of
1-forms) we will often use the notation
$$
\eptwo = e^\mu{}_a e^\nu{}_b - e^\mu{}_b e^\nu{}_a
$$
and similarly for $\epthree$.

\section{Massless case}

In this section we present a couple of most simple examples of
interacting massless theories in $AdS_3$. The reasons for its
inclusion are twofold. From one hand they give simple illustration of
general technics that we will use for the massive cases in the
subsequent sections. From the other one, it is instructive to compare
massless and massive cases because in some aspects they appear to be
drastically different.

\subsection{Gravity}

Free massless spin-2 filed in $AdS_3$ space is described by the
Lagrangian:
\begin{equation}
{\cal L}_0 = \frac{1}{2} \eptwo \omega_\mu{}^a \omega_\mu{}^b -
\varepsilon^{\mu\nu\alpha} \omega_\mu{}^a D_\nu h_\alpha{}^a +
\frac{\lambda^2}{2} \eptwo h_\mu{}^a h_\nu{}^b
\end{equation}
which is invariant under the following gauge transformations:
\begin{equation}
\delta_0 \omega_\mu{}^a = D_\mu \eta^a + \lambda^2
\varepsilon_\mu{}^{ab} \xi^b, \qquad \delta_0 h_\mu{}^a =  D_\mu \xi^a
+ \varepsilon_\mu{}^{ab}
\eta^b
\end{equation}
Now if we introduce two gauge invariant objects (curvature and
torsion)
\begin{eqnarray}
R_{\mu\nu}{}^a &=& D_{\mu} \omega_{\nu]}{}^a + \lambda^2
\varepsilon_{[\mu}{}^{ab} h_{\nu]}{}^b \nonumber \\
T_{\mu\nu}{}^a &=& D_{[\mu} h_{\nu]}{}^a + \varepsilon_{[\mu}{}^{ab}
\omega_{\nu]}{}^b
\end{eqnarray}
then the Lagrangian can be rewritten as follows
\begin{equation}
{\cal L}_0 = - \frac{1}{4} \varepsilon^{\mu\nu\alpha} [ \omega_\mu{}^a
T_{\nu\alpha}{}^a + h_\mu{}^a R_{\nu\alpha}{}^a ]
\end{equation}
Let us introduce new variables
\begin{equation}
\hat{\omega}_\mu{}^a = \omega_\mu{}^a + \lambda h_\mu{}^a, \qquad
\hat{h}_\mu{}^a = \omega_\mu{}^a - \lambda h_\mu{}^a
\end{equation}
and their corresponding gauge invariant objects
\begin{equation}
\hat{R}_{\mu\nu}{}^a = D_{[\mu} \hat{\omega}_{\nu]}{}^a + \lambda
\varepsilon_{[\mu}{}^{ab} \hat{\omega}_{\nu]}{}^b, \qquad
\hat{T}_{\mu\nu}{}^a = D_{[\mu} \hat{h}_{\nu]}{}^a - \lambda
\varepsilon_{[\mu}{}^{ab} \hat{h}_{\nu]}{}^b
\end{equation}
Then the Lagrangian takes the form
\begin{equation}
{\cal L}_0 = - \frac{1}{8\lambda} \varepsilon^{\mu\nu\alpha} [
\hat{\omega}_\mu{}^a \hat{R}_{\nu\alpha}{}^a - \hat{h}_\mu{}^a
\hat{T}_{\nu\alpha}{}^a ]
\end{equation}
Thus the Lagrangian now consists of two independent parts each one
containing only one field and is invariant under its own gauge
transformation
\begin{equation}
\delta_0 \hat{\omega}_\mu{}^a = D_\mu \hat{\eta}^a + \lambda
\varepsilon_\mu{}^{ab} \hat{\eta}^b, \qquad
\delta_0 \hat{h}_\mu{}^a = D_\mu \hat{\xi}^a - \lambda
\varepsilon_\mu{}^{ab} \hat{\xi}^b
\end{equation}
where
\begin{equation}
\hat{\eta} = \eta^a + \lambda \xi^a, \qquad
\hat{\xi}^a = \eta^a - \lambda \xi^a
\end{equation}
Now if we suppose that such possibility to separate variables exists
after switching on interaction as well we can greatly simplify all
calculations working with only one field and one gauge transformation.
In components the free Lagrangian for the field $\hat{\omega}_\mu{}^a$
has the form:
\begin{equation}
{\cal L}_0 = - \frac{1}{4\lambda} [ \varepsilon^{\mu\nu\alpha}
\hat{\omega}_\mu{}^a D_\nu \hat{\omega}_\alpha{}^a - \lambda \eptwo
\hat{\omega}_\mu{}^a \hat{\omega}_\nu{}^b ]
\end{equation}
To consider possible self interactions we begin with linear
approximation (i.e. cubic terms in the Lagrangian and linear terms in
gauge transformations). In this case the only possible cubic vertex
and corresponding correction to gauge transformations  look like:
\begin{equation}
{\cal L}_1 = a_0 \epthree \hat{\omega}_\mu{}^a \hat{\omega}_\nu{}^b
\hat{\omega}_\alpha{}^c, \qquad
\delta_1 \hat{\omega}_\mu{}^a = \alpha_0 \varepsilon^{abc}
\hat{\omega}_\mu{}^b \hat{\eta}^c
\end{equation}
Gauge invariance in the linear approximation requires that
$$
\delta_0 {\cal L}_1 + \delta_1 {\cal L}_0 =0
$$
and gives us
\begin{equation}
a_0 = - \frac{\alpha_0}{12\lambda}
\end{equation}
Now if we try to go beyond linear approximation we face the important
fact that in the frame-like formalism in $d=3$ there are no any
quartic vertices for all spins $s \ge 2$. Happily, in this particular
case it is easy to check that $\delta_1 {\cal L}_1 = 0$ so that we
have closed consistent theory without any need to introduce some other
fields. The resulting Lagrangian
\begin{equation}
{\cal L} = - \frac{1}{4\lambda} [ \varepsilon^{\mu\nu\alpha}
\hat{\omega}_\mu{}^a D_\nu \hat{\omega}_\alpha{}^a - \lambda \eptwo
\hat{\omega}_\mu{}^a \hat{\omega}_\nu{}^b + \frac{\alpha_0}{3}
\epthree \hat{\omega}_\mu{}^a \hat{\omega}_\nu{}^b
\hat{\omega}_\alpha{}^c ]
\end{equation}
is invariant under the following gauge transformations
\begin{equation}
\delta_0 \hat{\omega}_\mu{}^a = D_\mu \hat{\eta}^a + \lambda
\varepsilon_\mu{}^{ab} \hat{\eta}^b + \alpha_0 \varepsilon^{abc}
\hat{\omega}_\mu{}^b \hat{\eta}^c
\end{equation}
In turn, this gauge invariance implies that there exists deformation
for curvature
\begin{equation}
\hat{\tilde{R}}_{\mu\nu}{}^a = \hat{R}_{\mu\nu}{}^a +
\frac{\alpha_0}{2} \varepsilon^{abc} \hat{\omega}_{[\mu}{}^b
\hat{\omega}_{\nu]}{}^c
\end{equation}
which transforms covariantly
\begin{equation}
\delta \hat{\tilde{R}}_{\mu\nu}{}^a = \alpha_0 \varepsilon^{abc}
\hat{\tilde{R}}_{\mu\nu}{}^b \hat{\eta}^c
\end{equation}
In-particular, this means that this model can be considered as
Chern-Simons theory with the algebra $O(2,1) \approx SL(2)$
\cite{Wit88}. Note however that the interacting Lagrangian can not be
written in the form similar to the free one because
$$
\frac{1}{2} \varepsilon^{\mu\nu\alpha} \hat{\omega}_\mu{}^a
\hat{\tilde{R}}_{\nu\alpha}{}^a = \varepsilon^{\mu\nu\alpha}
\hat{\omega}_\mu{}^a D_\nu \hat{\omega}_\alpha{}^a - \lambda \eptwo
\hat{\omega}_\mu{}^a \hat{\omega}_\nu{}^b + \frac{\alpha_0}{2}
\epthree \hat{\omega}_\mu{}^a \hat{\omega}_\nu{}^b
\hat{\omega}_\alpha{}^c
$$
but the equations of motion following from the Lagrangian are
equivalent to
$$
\hat{\tilde{R}}_{\mu\nu}{}^a = 0
$$
Now we can easily return back to initial variables taking the complete
Lagrangian in the form
\begin{equation}
{\cal L} = {\cal L}(\hat{\omega}) - {\cal L}(\hat{h})
\end{equation}
Note that in terms of initial variables each part contains parity odd
higher derivative terms and only when coefficients are equal these
terms are canceled and we obtain parity even theory with no more than
two derivatives. Such theory is nothing else but usual $d=3$ gravity
in
$AdS$ background with the Lagrangian
\begin{eqnarray}
{\cal L} &=& \frac{1}{2} \eptwo \omega_\mu{}^a \omega_\mu{}^b -
\varepsilon^{\mu\nu\alpha} \omega_\mu{}^a D_\nu h_\alpha{}^a +
\frac{\lambda^2}{2} \eptwo h_\mu{}^a h_\nu{}^b - \nonumber \\
 && - \frac{\alpha_0}{2} \epthree [ \omega_\mu{}^a \omega_\nu{}^b
h_\alpha{}^c + \frac{\lambda^2}{3} h_\mu{}^a h_\nu{}^b h_\alpha{}^c ]
\end{eqnarray}
which is invariant under the following gauge transformations:
\begin{eqnarray}
\delta \omega_\mu{}^a &=& D_\mu \eta^a + \lambda^2
\varepsilon_\mu{}^{ab} \xi^b  + \alpha_0 \varepsilon^{abc} [
\omega_\mu{}^b \eta^c + \lambda^2 h_\mu{}^b \xi^c ]  \nonumber \\
\delta h_\mu{}^a &=& D_\mu \xi^a + \varepsilon_\mu{}^{ab} \eta^b +
\alpha_0 \varepsilon^{abc} [ h_\mu{}^b \eta^c + \omega_\mu{}^b \xi^c ]
\end{eqnarray}
At the same time in this formulation it is equivalent to Chern-Simons
theory with the algebra $O(2,2) \approx O(2,1) \times O(1,2) \approx
SL(2) \times SL(2)$ \cite{Wit88,CFPT10}. Note here that it is the
separation of variables that we use to simplify construction that
requires non-zero $\lambda$. As can be clearly seen from the last two
formulas in the complete theory nothing prevent us from considering
the flat limit $\lambda \to 0$. The resulting theory (it is just a
usual $d=3$ gravity in a flat Minkowski background) can still be
considered as a Chern-Simons one but with the algebra isomorphic to
three-dimensional Poincare one.

\subsection{Spin-3 field}

Free massless spin-3 field in $AdS_3$ space can be described by the
following Lagrangian
\begin{equation}
{\cal L}_0 = - \eptwo \Omega_\mu{}^{ac} \Omega_\nu{}^{bc} +
\varepsilon^{\mu\nu\alpha} \Omega_\mu{}^{ab} D_\nu \Phi_\alpha{}^{ab}
- \lambda^2 \eptwo \Phi_\mu{}^{ac} \Phi_\nu{}^{bc}
\end{equation}
where both $\Omega_\mu{}^{ab}$ and $\Phi_\mu{}^{ab}$ are symmetric and
traceless on local indices. This Lagrangian is invariant under the
following gauge transformations:
\begin{equation}
\delta_0 \Omega_\mu{}^{ab} = D_\mu \eta^{ab} - \lambda^2
\varepsilon_\mu{}^{c(a} \xi^{b)c}, \qquad
\delta_0 \Phi_\mu{}^{ab} = D_\mu \xi^{ab} - \varepsilon_\mu{}^{c(a}
\eta^{b)c}
\end{equation}
where both $\eta^{ab}$ and $\xi^{ab}$ are symmetric and traceless. As
in the spin 2 case the Lagrangian can be written in terms of gauge
invariant objects (which we will call curvatures)
\begin{equation}
{\cal G}_{\mu\nu}{}^{ab} = D_{[\mu} \Omega_{\nu]}{}^{ab} - \lambda^2
\varepsilon_{[\mu}{}^{c(a} \Phi_{\nu]}{}^{b)c}, \qquad
{\cal H}_{\mu\nu}{}^{ab} = D_{[\mu} \Phi_{\nu]}{}^{ab} -
\varepsilon_{[\mu}{}^{c(a} \Omega_{\nu]}{}^{b)c}
\end{equation}
in the form
\begin{equation}
{\cal L}_0 = \frac{1}{4} \varepsilon^{\mu\nu\alpha} [
\Omega_\mu{}^{ab} {\cal H}_{\nu\alpha}{}^{ab} + \Phi_\mu{}^{ab}
{\cal G}_{\nu\alpha}{}^{ab} ]
\end{equation}
As in the spin-2 case we can introduce new variables
\begin{equation}
\hat{\Omega}_\mu{}^{ab} = \Omega_\mu{}^{ab} + \lambda \Phi_\mu{}^{ab},
\qquad \hat{\Phi}_\mu{}^{ab} = \Omega_\mu{}^{ab} - \lambda
\Phi_\mu{}^{ab}
\end{equation}
and rewrite the free Lagrangian in the form
\begin{equation}
{\cal L}_0 = \frac{1}{8\lambda} \varepsilon^{\mu\nu\alpha} [
\hat{\Omega}_\mu{}^{ab} \hat{\cal G}_{\nu\alpha}{}^{ab} -
\hat{\Phi}_\mu{}^{ab} \hat{\cal H}_{\nu\alpha}{}^{ab} ]
\end{equation}
where
\begin{equation}
\hat{\cal G}_{\mu\nu}{}^{ab} = D_{[\mu} \hat{\Omega}_{\nu]}{}^{ab} -
\lambda \varepsilon_{[\mu}{}^{c(a} \hat{\Omega}_{\nu]}{}^{b)c}, \qquad
\hat{\cal H}_{\mu\nu}{}^{ab} = D_{[\mu} \hat{\Phi}_{\nu]}{}^{ab} +
\lambda \varepsilon_{[\mu}{}^{c(a} \hat{\Phi}_{\nu]}{}^{b)c}
\end{equation}
Again each half of the Lagrangian contains one field only and is
invariant under one gauge transformation
\begin{equation}
\delta_0 \hat{\Omega}_\mu{}^{ab} = D_\mu \hat{\eta}^{ab} - \lambda
\varepsilon_\mu{}^{c(a} \hat{\eta}^{b)c}, \qquad
\delta_0 \hat{\Phi}_\mu{}^{ab} = D_\mu \hat{\xi}^{ab} + \lambda
\varepsilon_\mu{}^{c(a} \hat{\xi}^{b)c}
\end{equation}
where now
\begin{equation}
\hat{\eta}^{ab} = \eta^{ab} + \lambda \xi^{ab}. \qquad
\hat{\xi}^{ab} = \eta^{ab} - \lambda \xi^{ab}
\end{equation}
Thus assuming that the possibility to separate variables exists after
switching on interaction we can work with one field and take care on
one gauge transformation. The component form for
$\hat{\Omega}_\mu{}^{ab}$ field Lagrangian looks as follows
\begin{equation}
{\cal L}_0 = \frac{1}{4\lambda} [ \varepsilon^{\mu\nu\alpha}
\hat{\Omega}_\mu{}^{ab} D_\nu \hat{\Omega}_\alpha{}^{ab} - 2\lambda
\eptwo \hat{\Omega}_\mu{}^{ac} \hat{\Omega}_\nu{}^{bc} ]
\end{equation}
Now let us turn to the gravitational interactions for such particles.
The only possible cubic vertex now has the form
\begin{equation}
{\cal L}_1 = a_1 \epthree \hat{\Omega}_\mu{}^{ad}
\hat{\Omega}_\nu{}^{bd} \hat{\omega}_\alpha{}^c
\end{equation}
while corresponding corrections to gauge transformations can be
written as follows
\begin{equation}
\delta_1 \hat{\Omega}_\mu{}^{ab} = \alpha_1 \varepsilon^{cd(a}
\hat{\Omega}_\mu{}^{b)c} \hat{\eta}^d + \alpha_2 \varepsilon^{cd(a}
\hat{\eta}^{b)c} \hat{\omega}_\mu{}^d, \qquad
\delta_1 \hat{\omega}_\mu{}^a = \alpha_3 \varepsilon^{abc}
\hat{\Omega}_\mu{}^{bd} \hat{\eta}^{cd}
\end{equation}
Now consider gauge invariance in the linear approximation that
requires $\delta_0 {\cal L}_1 + \delta_1 {\cal L}_0 = 0$. For
$\hat{\eta}^a$ transformations we obtain
\begin{eqnarray*}
\delta_0 {\cal L}_1 &=& - 2a_1 \epthree D_\mu \hat{\Omega}_\nu{}^{ad}
\hat{\Omega}_\nu{}^{bd} \hat{\eta}^c + 2a_1 \lambda
\varepsilon^{\mu\nu\alpha} \hat{\Omega}_{\mu,\nu}{}^a
\hat{\Omega}_\alpha{}^{ab} \hat{\eta}^b \\
\delta_1 {\cal L}_0 &=& \frac{\alpha_1}{\lambda} \epthree D_\mu
\hat{\Omega}_\nu{}^{ad} \hat{\Omega}_\nu{}^{bd} \hat{\eta}^c -
\alpha_1 \varepsilon^{\mu\nu\alpha} \hat{\Omega}_{\mu,\nu}{}^a
\hat{\Omega}_\alpha{}^{ab} \hat{\eta}^b
\end{eqnarray*}
This gives us
$$
a_1 = \frac{\alpha_1}{2\lambda}
$$
At  the same time for $\hat{\eta}^{ab}$ transformations we get
\begin{eqnarray*}
\delta_0 {\cal L}_1 &=& 2a_1 \epthree [ D_\mu \hat{\Omega}_\nu{}^{ad}
\hat{\eta}^{bd} \hat{\omega}_\alpha{}^c - \hat{\Omega}_\mu{}^{ad}
\hat{\eta}^{bd} D_\nu \hat{\omega}_\alpha{}^c ] + \\
 && + 2a_1\lambda \varepsilon^{\mu\nu\alpha} [ 2
\hat{\Omega}_{\mu,\nu}{}^a \hat{\eta}^{ab} \hat{\omega}_\alpha{}^c +
\hat{\Omega}_\mu{}^{ab} \hat{\eta}_\nu{}^a \hat{\omega}_\alpha{}^b + 2
\hat{\Omega}_\mu{}^{ab} \hat{\eta}{}^{ab} \hat{\omega}_{\nu,\alpha} ]
\\
\delta_1 {\cal L}_0 &=& \epthree [ \frac{\alpha_2}{\lambda} D_\mu
\hat{\Omega}_\nu{}^{ad} \hat{\eta}^{bd} \hat{\omega}_\alpha{}^c -
\frac{\alpha_3}{2\lambda} \hat{\Omega}_\mu{}^{ad} \hat{\eta}^{bd}
D_\nu \hat{\omega}_\alpha{}^c ] + \\
 && + \varepsilon^{\mu\nu\alpha} [ (\alpha_2 + \frac{\alpha_3}{2}
\hat{\Omega}_{\mu,\nu}{}^a \hat{\eta}^{ab} \hat{\omega}_\alpha{}^c +
(2\alpha_2 - \frac{\alpha_3}{2}) \hat{\Omega}_\mu{}^{ab}
\hat{\eta}_\nu{}^a \hat{\omega}_\alpha{}^b + 2\alpha_2
\hat{\Omega}_\mu{}^{ab} \hat{\eta}{}^{ab} \hat{\omega}_{\nu,\alpha} ]
\end{eqnarray*}
and we obtain
$$
\alpha_2 = - \alpha_1, \qquad \alpha_3 = - 2 \alpha_1
$$
To go beyond linear approximation let us collect together self
interaction for graviton and its interaction with spin 3:
\begin{equation}
{\cal L}_1 = - \frac{\alpha_0}{12\lambda} \epthree
\hat{\omega}_\mu{}^a \hat{\omega}_\nu{}^b \hat{\omega}_\alpha{}^c +
\frac{\alpha_1}{2\lambda} \epthree \hat{\Omega}_\mu{}^{ad}
\hat{\Omega}_\nu{}^{bd} \hat{\omega}_\alpha{}^c
\end{equation}
and all corrections to gauge transformations:
\begin{eqnarray}
\delta_1 \hat{\omega}_\mu{}^a &=& \alpha_0 \varepsilon^{abc}
\hat{\omega}_\mu{}^b \hat{\eta}^c - 2\alpha_1 \varepsilon^{abc}
\hat{\Omega}_\nu{}^{bd} \hat{\eta}^{cd} \nonumber \\
\delta_1 \hat{\Omega}_\mu{}^{ab} &=& \alpha_1 \varepsilon^{cd(a}
\hat{\Omega}_\mu{}^{b)c} \hat{\eta}^d - \alpha_1 \varepsilon^{cd(a}
\hat{\eta}^{b)c} \hat{\omega}_\mu{}^d
\end{eqnarray}
Now calculating quadratic variations we obtain
$$
\delta_1 {\cal L}_1 = \frac{(\alpha_0-\alpha_1)\alpha_1}{\lambda}
\varepsilon^{\mu\nu\alpha} [ \hat{\Omega}_\mu{}^{ab}
\hat{\Omega}_\nu{}^{ac} \hat{\omega}_\alpha{}^b \hat{\eta}^c +
\hat{\omega}_\mu{}^a \hat{\omega}_\nu{}^b \hat{\Omega}_\alpha{}^{ac}
\hat{\eta}^{bc} ]
$$
so that for $\alpha_1 = \alpha_0$ all variations cancel. As in the
pure gravity case the invariance of the Lagrangian implies that there
exist deformations of the curvatures
\begin{eqnarray}
\hat{\tilde{R}}_{\mu\nu}{}^a &=& \hat{R}_{\mu\nu}{}^a +
\frac{\alpha_0}{2} \varepsilon^{abc} \hat{\omega}_{[\mu}{}^b
\hat{\omega}_{\nu]}{}^c - \alpha_0 \varepsilon^{abc}
\hat{\Omega}_{[\mu}{}^{bd} \hat{\Omega}_{\nu]}{}^{cd} \nonumber \\
\hat{\tilde{\cal G}}_{\mu\nu}{}^{ab} &=& \hat{\cal G}_{\mu\nu}{}^{ab}
+ \alpha_0 \varepsilon^{cd(a} \hat{\Omega}_{[\mu}{}^{b)c}
\hat{\omega}_{\nu]}{}^d
\end{eqnarray}
which transform covariantly
\begin{eqnarray}
\delta \hat{\tilde{R}}_{\mu\nu}{}^a &=& \alpha_0 \varepsilon^{abc} [
\hat{\tilde{R}}_{\mu\nu}{}^b \hat{\eta}^c - 2
\hat{\tilde{\cal G}}_{\mu\nu}{}^{bd} \hat{\eta}^{cd} ] \nonumber \\
\delta \hat{\tilde{\cal G}}_{\mu\nu}{}^{ab} &=& \alpha_0
\varepsilon^{cd(a} [ \hat{\tilde{\cal G}}_{\mu\nu}{}^{b)c}
\hat{\eta}^d - \hat{\eta}^{b)c} \hat{\tilde{R}}_{\mu\nu}{}^d ]
\end{eqnarray}
and such that the equations of motion following from the Lagrangian
are equivalent to
\begin{equation}
\hat{\tilde{R}}_{\mu\nu}{}^a = 0, \qquad
\hat{\tilde{\cal G}}_{\mu\nu}{}^{ab} = 0
\end{equation}
In this, such model can be considered as a Chern-Simons theory with
the algebra $SL(3)$ \cite{CFPT10}.

To simplify comparison with the massive case let us re-express main
formulas in terms of the initial variables. The interacting Lagrangian
has the form:
\begin{eqnarray}
{\cal L}_1 &=& \alpha_0 \epthree [ \Omega_\mu{}^{ad} \Omega_\nu{}^{bd}
h_\alpha{}^c + 2 \Omega_\mu{}^{ad} \Phi_\nu{}^{bd} \omega_\alpha{}^c +
2\lambda^2 \Phi_\mu{}^{ad} \Phi_\nu{}^{bd} h_\alpha{}^c - \nonumber \\
 && \qquad \qquad - \frac{1}{2} \omega_\mu{}^a \omega_\nu{}^b
h_\alpha{}^c - \frac{\lambda^2}{6} h_\mu{}^a h_\nu{}^b h_\alpha{}^c ]
\end{eqnarray}
while corrections to gauge transformations look as follows:
\begin{eqnarray}
\delta_1 \omega_\mu{}^a &=& \alpha_0 \varepsilon^{abc} [
\omega_\mu{}^b \eta^c + \lambda^2 h_\mu{}^b \xi^c - 2
\Omega_\mu{}^{bd} \eta^{cd} - 2\lambda^2 \Phi_\mu{}^{bd} \xi^{cd} ]
\nonumber \\
\delta_1 h_\mu{}^a &=& \alpha_0 \varepsilon^{abc} [ h_\mu{}^b \eta^c +
\omega_\mu{}^b \xi^c - 2 \Phi_\mu{}^{bd} \eta^{cd} - 2
\Omega_\mu{}^{bd} \xi^{cd} ] \nonumber \\
\delta_1 \Omega_\mu{}^{ab} &=& \alpha_0 \varepsilon^{cd(a} [
\Omega_\mu{}^{b)c} \eta^d + \lambda^2 \Phi_\mu{}^{b)c} \xi^d -
\eta^{b)c} \omega_\mu{}^d - \lambda^2 \xi^{b)c} h_\mu{}^d ] \\
\delta_1 \Phi_\mu{}^{ab} &=& \alpha_0 \varepsilon^{cd(a} [
\Phi_\mu{}^{b)c} \eta^d + \Omega_\mu{}^{b)c} \xi^d - \eta^{b)c}
h_\mu{}^d - \xi^{b)c} \omega_\mu{}^d ]   \nonumber
\end{eqnarray}
At last the deformed curvatures can be written as:
\begin{eqnarray}
\tilde{R}_{\mu\nu}{}^a &=& R_{\mu\nu}{}^a + \frac{\alpha_0}{2}
\varepsilon^{abc} [ \omega_{[\mu}{}^b \omega_{\nu]}{}^c + \lambda^2
h_{[\mu}{}^b h_{\nu]}{}^c - 2 \Omega_{[\mu}{}^{bd}
\Omega_{\nu]}{}^{cd} - 2\lambda^2 \Phi_{[\mu}{}^{bd}
\Phi_{\nu]}{}^{cd} ]  \nonumber \\
\tilde{T}_{\mu\nu}{}^a &=& T_{\mu\nu}{}^a + \alpha_0 \varepsilon^{abc}
[ \omega_{[\mu}{}^b h_{\nu]}{}^c - 2 \Omega_{[\mu}{}^{bd}
\Phi_{\nu]}{}^{cd} ] \nonumber  \\
\tilde{\cal G}_{\mu\nu}{}^{ab} &=& {\cal G}_{\mu\nu}{}^{ab} + \alpha_0
\varepsilon^{cd(a} [ \Omega_{[\mu}{}^{b)c} \omega_{\nu]}{}^d +
\lambda^2 \Phi_{[\mu}{}^{b)c} h_{\nu]}{}^d ]  \\
\tilde{\cal H}_{\mu\nu}{}^{ab} &=& {\cal H}_{\mu\nu}{}^{ab} + \alpha_0
\varepsilon^{cd(a} [ \Phi_{[\mu}{}^{b)c} \omega_{\nu]}{}^d +
\Omega_{[\mu}{}^{b)c} h_{\nu]}{}^d ] \nonumber
\end{eqnarray}
Thus complete model in terms of initial variables is equivalent to
Chern-Simons theory with the algebra $SL(3) \times SL(3)$
\cite{CFPT10}. Let us stress here that contrary to higher-dimensional
case \cite{FV87,FV87a} this model does not contain any higher
derivative terms and as a result admits non-singular flat limit
$\lambda \to 0$.

\subsection{Arbitrary spin-$s$ field}

For brevity in this case from the very beginning we will work in
terms of separated variable $\hat{\Omega}_\mu{}^{a_1 \dots a_{s-1}}$
which are completely symmetric and traceless on local indices. Also
to simplify formulas we will use compact notations
$\Omega_\mu{}^{a_1 \dots a_k} = \Omega_\mu{}^{(k)}$ where index $k$
denotes just the number of local indices and not the indices
themselves. The free Lagrangian for massless spin-$s$ field has the
form \cite{BSZ12a}:
\begin{equation}
{\cal L}_0 = \frac{(-1)^{s+1}}{4\lambda} [ \varepsilon^{\mu\nu\alpha}
\hat{\Omega}_\mu{}^{(s-1)} D_\nu \hat{\Omega}_\alpha{}^{(s-1)} -
(s-1)\lambda \eptwo \hat{\Omega}_\mu{}^{a(s-2)}
\hat{\Omega}_\nu{}^{b(s-2)} ]
\end{equation}
It is invariant under the following gauge transformations
\begin{equation}
\delta_0 \hat{\Omega}_\mu{}^{(s-1)} = D_\mu \hat{\eta}^{(s-1)} -
\varepsilon_\mu{}^{a(1} \hat{\eta}^{s-2)a}
\end{equation}
where parameter $\hat{\eta}^{(s-1)}$ is also symmetric and traceless.

Let us consider interaction of this field with gravity. The only cubic
vertex possible looks like:
\begin{equation}
{\cal L}_1 = (-1)^{s+1} a_s \epthree \hat{\Omega}_\mu{}^{a(s-2)}
\hat{\Omega}_\nu{}^{b(s-2)} \hat{\omega}_\alpha{}^c
\end{equation}
while corresponding corrections to gauge transformations can be
written as follows:
\begin{eqnarray}
\delta_1 \hat{\Omega}_\mu{}^{(s-1)} &=& \alpha_1 \varepsilon^{ab(1}
\hat{\Omega}_\mu{}^{s-2)a} \hat{\eta}^b + \alpha_2 \varepsilon^{ab(1}
\hat{\eta}^{s-2)a} \hat{\omega}_\mu{}^b \nonumber \\
\delta_1 \hat{\omega}_\mu{}^a &=& \alpha_3 \varepsilon^{abc}
\hat{\Omega}_\mu{}^{b(s-2)} \hat{\eta}^{c(s-2)}
\end{eqnarray}
Now we have to calculate linear variations. For $\hat{\eta}^a$
transformations we obtain:
\begin{eqnarray*}
\delta_0 {\cal L}_1 &=& - 2a_s \epthree D_\mu
\hat{\Omega}_\nu{}^{a(s-2)} \hat{\Omega}_\alpha{}^{b(s-2)}
\hat{\eta}^c + 2 a_s \lambda \varepsilon^{\mu\nu\alpha}
\hat{\Omega}_{\mu,\nu}{}^{(s-2)} \hat{\Omega}_\alpha{}^{a(s-2)}
\hat{\eta}^a  \\
\delta_1 {\cal L}_0 &=& \frac{(s-1)\alpha_1}{2\lambda} [ \epthree
D_\mu \hat{\Omega}_\nu{}^{a(s-2)} \hat{\Omega}_\alpha{}^{b(s-2)}
\hat{\eta}^c - \lambda \varepsilon^{\mu\nu\alpha}
\hat{\Omega}_{\mu,\nu}{}^{(s-2)} \hat{\Omega}_\alpha{}^{a(s-2)}
\hat{\eta}^a ]
\end{eqnarray*}
Thus we have to put
$$
a_s = \frac{(s-1)\alpha_1}{4\lambda}
$$
Similarly, from the variations under the $\hat{\eta}^{(s-1)}$
transformations we get
$$
\alpha_2 = (-1)^s \alpha_1, \qquad \alpha_3 = - (s-1) \alpha_1
$$
Thus we have achieved gauge invariance in the linear approximation.
Combining self interaction for graviton with its interaction with spin
$s$ particle we obtain the cubic Lagrangian
\begin{equation}
{\cal L}_1 = \frac{1}{4\lambda} \epthree [ - \frac{\alpha_0}{3}
\hat{\omega}_\mu{}^a \hat{\omega}_\nu{}^b \hat{\omega}_\alpha{}^c +
(-1)^{s+1} (s-1) \alpha_1 \hat{\Omega}_\mu{}^{a(s-2)}
\hat{\Omega}_\nu{}^{b(s-2)} \hat{\omega}_\alpha{}^c ]
\end{equation}
and complete set of corrections to gauge transformations
\begin{eqnarray}
\delta_1 \hat{\omega}_\mu{}^a &=& \alpha_0 \varepsilon^{abc}
\hat{\omega}_\mu{}^b \hat{\eta}^c - (s-1) \alpha_1 \varepsilon^{abc}
\hat{\Omega}_\mu{}^{b(s-2)} \hat{\eta}^{c(s-2)} \nonumber \\
\delta_1 \hat{\Omega}_\mu{}^{(s-1)} &=& \alpha_1 \varepsilon^{ab(1}
\hat{\Omega}_\mu{}^{s-2)a} \hat{\eta}^b + (-1)^s \alpha_1
\varepsilon^{ab(1} \hat{\eta}^{s-2)a} \hat{\omega}_\mu{}^b
\end{eqnarray}
Now let us consider quadratic variations. For the $\hat{\eta}^a$
transformations we get
$$
\delta_1 {\cal L}_1 = \frac{(-1)^{s+1} (s-1) \alpha_1
(\alpha_0-\alpha_1)}{2\lambda} \varepsilon^{\mu\nu\alpha}
\hat{\Omega}_\mu{}^{a(s-2)} \hat{\Omega}_\nu{}^{b(s-2)}
\hat{\omega}_\alpha{}^a \hat{\eta}^b
$$
so we have to put $\alpha_1 = \alpha_0$. But this still leaves us with
the $\hat{\eta}^{(s-1)}$ variations of the form
$$
\delta_1 {\cal L}_1 \sim \hat{\Omega}^{(s-1)} \hat{\Omega}^{(s-1)}
\hat{\Omega}^{(s-1)} \hat{\eta}^{(s-1)}
$$
that cancel for the $s=3$ case only so that for any spin $s \ge 4$ to
obtain closed consistent theory we have to introduce some other fields
and/or interactions. As it has been shown in \cite{CFPT10} one of the
possible solutions is to introduce all intermediate spins
$s=3,4,\dots,(s-1)$ as well as corresponding additional cubic vertices
with the result being the Chern-Simons theory with the algebra
$SL(s)$.

\section{Massive spin-3 field}

Self interaction and interacting with gravity for massive spin-2
field has been considered in the recent paper of one of us
\cite{Zin12} and in this section we consider gravitational
interaction for massive spin 3 one.

\subsection{Kinematics of massive spin-3 field}

Frame-like gauge invariant formalism for massive spin-3 field in
$(A)dS$ space adopted to three dimensions \cite{BSZ12a} requires
four pairs of fields $(\Omega_\mu{}^{ab},\Phi_\mu{}^{ab})$,
$(\Omega_\mu{}^a,f_\mu{}^a)$, $(B^a, A_\mu)$ and $(\pi^a,\varphi)$.
In this, the free Lagrangian has the form
\begin{eqnarray}
{\cal{L}}_0 &=& - \eptwo \Omega_\mu{}^{ac} \Omega_\nu{}^{bc} +
\varepsilon^{\mu\nu\alpha} \Omega_\mu{}^{ab} D_\nu \Phi_\alpha{}^{ab}
+ \frac12 \eptwo \Omega_\mu{}^a \Omega_\nu{}^b -
\varepsilon^{\mu\nu\alpha} \Omega_\mu{}^a D_\nu f_\alpha{}^a +
\nonumber \\
&& + \frac12 B^a B^a - \varepsilon^{\mu\nu\alpha} B_\mu D_\nu
A_\alpha - \frac12 \pi^a \pi^a + \pi^\mu D_\mu \varphi + \nonumber\\
&& + \varepsilon^{\mu\nu\alpha}[ 3m \Omega_{\mu,\nu}{}^a f_\alpha{}^a
+ m \Phi_{\mu,\nu}{}^a \Omega_\alpha{}^a - 2\tilde{m} \Omega_{\mu,\nu}
A_\alpha + \tilde{m} f_{\mu,\nu} B_\alpha] + M \pi^\mu A_\mu +
\nonumber\\
&& + \eptwo [-\frac{M^2}{36} \Phi_\mu{}^{ac} \Phi_\nu{}^{bc} +
\frac{M^2}{8} f_\mu{}^a f_\nu{}^b] + M\tilde{m} e^\mu{}_a f_\mu{}^a
\varphi + 3\tilde{m}^2 \varphi^2 \label{s3l0}
\end{eqnarray}
and is invariant under the following set of gauge transformations
\begin{eqnarray}
\delta_0 \Omega_\mu{}^{ab} &=& D_\mu \eta^{ab} +
\frac{m}{2}(e_\mu{}^{(a} \eta^{b)} - \frac23 g^{ab} \eta_\mu) -
\frac{M^2}{36} \varepsilon_\mu{}^{c(a} \xi^{b)c} \nonumber \\
\delta_0 \Phi_\mu{}^{ab} &=& D_\mu \xi^{ab} - \varepsilon_\mu{}^{c(a}
\eta^{b)c} + \frac{3m}{2}(e_\mu{}^{(a} \xi^{b)} - \frac23 g^{ab}
\xi_\mu) \nonumber \\
\delta_0 \Omega_\mu{}^a &=& D_\mu \eta^a + 3m \eta_\mu{}^a +
\frac{M^2}{4} \varepsilon_\mu{}^{ab} \xi^b \label{s3d0} \\
\delta_0 f_\mu{}^a &=& D_\mu \xi^a + \varepsilon_\mu{}^{ab} \eta^b + m
\xi_\mu{}^a + 2\tilde{m} e_\mu{}^a \xi \nonumber \\
\delta_0 B^a &=& - 2\tilde{m} \eta^a, \qquad
\delta_0 A_\mu = D_\mu \xi + \tilde{m} \xi_\mu \nonumber \\
\delta_0 \pi^a &=& M\tilde{m} \xi^a, \qquad
\delta_0 \varphi = - M \xi \nonumber
\end{eqnarray}
where
$$
\tilde{m}^2 = 8m^2 + 4\lambda^2,
\qquad M^2 = 18(3m^2 + 2\lambda^2)
$$

Recall that in de Sitter space ($\lambda^2 < 0$) there exist  two
partially massless limits \cite{DW01a,DW01c,Zin01,Zin08b}. First one
corresponds to $M \to 0$, in this scalar component decouples leaving
us with the Lagrangian
\begin{eqnarray}
{\cal{L}}_0 &=& - \eptwo \Omega_\mu{}^{ac} \Omega_\nu{}^{bc} +
\varepsilon^{\mu\nu\alpha} \Omega_\mu{}^{ab} D_\nu \Phi_\alpha{}^{ab}
+ \frac12 \eptwo \Omega_\mu{}^a \Omega_\nu{}^b -
\varepsilon^{\mu\nu\alpha} \Omega_\mu{}^a D_\nu f_\alpha{}^a +
\nonumber \\
&& + \frac12 B^a B^a - \varepsilon^{\mu\nu\alpha} B_\mu D_\nu
A_\alpha + \nonumber\\
&& + \varepsilon^{\mu\nu\alpha}[ 3m \Omega_{\mu,\nu}{}^a f_\alpha{}^a
+ m \Phi_{\mu,\nu}{}^a \Omega_\alpha{}^a - 2\tilde{m} \Omega_{\mu,\nu}
A_\alpha + \tilde{m} f_{\mu,\nu} B_\alpha]
\end{eqnarray}
which is still invariant under the whole set of gauge transformations
\begin{eqnarray}
\delta_0 \Omega_\mu{}^{ab} &=& D_\mu \eta^{ab} +
\frac{m}{2}(e_\mu{}^{(a} \eta^{b)} - \frac23 g^{ab} \eta_\mu)
\nonumber \\
\delta_0 \Phi_\mu{}^{ab} &=& D_\mu \xi^{ab} - \varepsilon_\mu{}^{c(a}
\eta^{b)c} + \frac{3m}{2}(e_\mu{}^{(a} \xi^{b)} - \frac23 g^{ab}
\xi_\mu) \nonumber \\
\delta_0 \Omega_\mu{}^a &=& D_\mu \eta^a + 3m \eta_\mu{}^a \\
\delta_0 f_\mu{}^a &=& D_\mu \xi^a + \varepsilon_\mu{}^{ab} \eta^b + m
\xi_\mu{}^a + 2\tilde{m} e_\mu{}^a \xi \nonumber \\
\delta_0 B^a &=& - 2\tilde{m} \eta^a, \qquad
\delta_0 A_\mu = D_\mu \xi + \tilde{m} \xi_\mu \nonumber
\end{eqnarray}
In $d=3$ dimensions such partially massless particle has one physical
degree of freedom only instead of usual two in general massive case.

The other partially massless limit corresponds to $\tilde{m} \to 0$,
in this scalar and vector components decouple leaving us with the
Lagrangian
\begin{eqnarray}
{\cal{L}}_0 &=& - \eptwo \Omega_\mu{}^{ac} \Omega_\nu{}^{bc} +
\varepsilon^{\mu\nu\alpha} \Omega_\mu{}^{ab} D_\nu \Phi_\alpha{}^{ab}
+ \frac12 \eptwo \Omega_\mu{}^a \Omega_\nu{}^b -
\varepsilon^{\mu\nu\alpha} \Omega_\mu{}^a D_\nu f_\alpha{}^a +
\nonumber \\
&& + \varepsilon^{\mu\nu\alpha}[ 3m \Omega_{\mu,\nu}{}^a f_\alpha{}^a
+ m \Phi_{\mu,\nu}{}^a \Omega_\alpha{}^a ] + \eptwo [-\frac{M^2}{36}
\Phi_\mu{}^{ac} \Phi_\nu{}^{bc} + \frac{M^2}{8} f_\mu{}^a f_\nu{}^b]
\end{eqnarray}
which is invariant under the following gauge transformations:
\begin{eqnarray}
\delta_0 \Omega_\mu{}^{ab} &=& D_\mu \eta^{ab} +
\frac{m}{2}(e_\mu{}^{(a} \eta^{b)} - \frac23 g^{ab} \eta_\mu) -
\frac{M^2}{36} \varepsilon_\mu{}^{c(a} \xi^{b)c} \nonumber \\
\delta_0 \Phi_\mu{}^{ab} &=& D_\mu \xi^{ab} - \varepsilon_\mu{}^{c(a}
\eta^{b)c} + \frac{3m}{2}(e_\mu{}^{(a} \xi^{b)} - \frac23 g^{ab}
\xi_\mu) \nonumber \\
\delta_0 \Omega_\mu{}^a &=& D_\mu \eta^a + 3m \eta_\mu{}^a +
\frac{M^2}{4} \varepsilon_\mu{}^{ab} \xi^b \\
\delta_0 f_\mu{}^a &=& D_\mu \xi^a + \varepsilon_\mu{}^{ab} \eta^b + m
\xi_\mu{}^a  \nonumber
\end{eqnarray}
In such limit we obtain a system that does not have any physical
degrees of freedom. At last, recall that in the massless limit in
$AdS$ space massive spin 3 field decomposes into massless spin 3
field and massive spin-2 one.

Let us return back to general massive case. Similarly to the
massless case for all fields entering the description of massive
field one can construct corresponding gauge invariant object (which
we will call curvatures though there are two forms as well as one
forms among them):
\begin{eqnarray}
{\cal G}_{\mu\mu}{}^{ab} &=& D_{[\mu} \Omega_{\nu]}{}^{ab} +
\frac{m}{2} ( e_{[\mu}{}^{(a} \Omega_{\nu]}{}^{b)} + \frac{2}{3}
g^{ab} \Omega_{[\mu,\nu]} ) - \frac{M^2}{36}
\varepsilon_{[\mu}{}^{c(a} \Phi_{\nu]}{}^{b)c} \nonumber \\
{\cal H}_{\mu\nu}{}^{ab} &=& D_{[\mu} \Phi_{\nu]}{}^{ab} -
\varepsilon_{[\mu}{}^{c(a} \Omega_{\nu]}{}^{b)c} + \frac{3m}{2} (
e_{[\mu}{}^{(a} f_{\nu]}{}^{b)} + \frac{2}{3} g^{ab} f_{[\mu,\nu]})
\nonumber \\
{\cal F}_{\mu\nu}{}^a &=& D_{[\mu} \Omega_{\nu]}{}^a - 3m
\Omega_{[\mu,\nu]}{}^a - \tilde{m} e_{[\mu}{}^a B_{\nu]} +
\frac{M^2}{4} \varepsilon_{[\mu}{}^{ab} f_{\nu]}{}^b - M\tilde{m}
\varepsilon_{\mu\nu}{}^a \varphi \nonumber \\
{\cal T}_{\mu\nu}{}^a &=& D_{[\mu} f_{\nu]}{}^a +
\varepsilon_{[\mu}{}^{ab} \Omega_{\nu]}{}^b - m \Phi_{[\mu,\nu]}{}^a +
2\tilde{m} e_{[\mu}{}^a A_{\nu]} \nonumber \\
{\cal B}_\mu{}^a &=& D_\mu B^a + 2\tilde{m} \Omega_\mu{}^a -
\frac{M}{2} \varepsilon_\mu{}^{ab} \pi^b - V_\mu{}^a \\
{\cal A}_{\mu\nu} &=& D_{[\mu} A_{\nu]} - \varepsilon_{\mu\nu}{}^a B^a
- \tilde{m} f_{[\mu,\nu]} \nonumber \\
\Pi_\mu{}^a &=& D_\mu \pi^a - M\tilde{m} f_\mu{}^a - 2\tilde{m}^2
e_\mu{}^a \varphi - W_\mu{}^a \nonumber \\
\Phi_\mu &=& D_\mu \varphi - \pi_\mu + M A_\mu \nonumber
\end{eqnarray}
where zero forms $V^{ab}$ and $W^{ab}$ (that do not enter the free
Lagrangian) are symmetric, traceless and have the following
transformations:
$$
\delta_0 V^{ab} = 6m\tilde{m} \eta^{ab}, \qquad
\delta_0 W^{ab} = - Mm\tilde{m} \xi^{ab}
$$
Moreover, using these curvatures the free Lagrangian can be rewritten
in the form similar to the Chern-Simons one
\begin{eqnarray}
{\cal L}_0 &=& \frac{1}{4} \varepsilon^{\mu\nu\alpha} [
\Omega_\mu{}^{ab} {\cal H}_{\nu\alpha}{}^{ab} + \Phi_\mu{}^{ab}
{\cal G}_{\nu\alpha}{}^{ab} - \Omega_\mu{}^a {\cal T}_{\nu\alpha}{}^a
- f_\mu{}^a {\cal F}_{\nu\alpha}{}^a - \nonumber \\
 && \qquad - 2 A_\mu {\cal B}_{\nu,\alpha} - B_\mu
{\cal A}_{\nu\alpha} ] + \frac{1}{2} e^\mu{}_a [ - \varphi \Pi_\mu{}^a
+ \pi^a \Phi_\mu ]
\end{eqnarray}

\subsection{Non-canonical vertices}

In our investigations of gravitational vertices we will call vertex
canonical if it corresponds to switching on standard minimal
interactions, i.e. to the replacement of background frame
$e_\mu{}^a$ by the dynamical one $e_\mu{}^a - h_\mu{}^a$ and the
background Lorentz derivative $D_\mu$ by fully covariant one $D_\mu
- \omega_\mu$. For massive spin-3 case due to the presence of
Stueckelberg fields we have two possible non-canonical vertices.
First one of the type $2-2-0$ where one of the spin 2 is the
graviton, while the other is a Stueckelberg field. Such vertex has
been considered in the work of one of us \cite{Zin12} devoted to
massive spin-2 where it was shown that such vertex can be completely
removed by field redefinitions. The second possibility is the vertex
of type $3-2-1$ where spin-2 is again the graviton, while spin 1 is
a Stueckelberg field. The most general ansatz for such vertex can be
written as follows:
\begin{eqnarray}
{\cal L}_1 &=& \eptwo [ a_1 \Omega_\mu{}^{ac} h_\nu{}^c B^b + a_2
\Omega_\mu{}^{ac} h_\nu{}^b B^c + a_3 \Phi_\mu{}^{ac} \omega_\nu{}^c
B^b + a_4 \Phi_\mu{}^{ac} \omega_\nu{}^b B^c ] + \nonumber \\
 && + \varepsilon^{\mu\nu\alpha} [ a_5 h_\mu{}^a D_\nu
\Phi_\alpha{}^{ab} B^b + a_6 D_\mu h_\nu{}^a \Phi_\alpha{}^{ab}
B^b ]
\end{eqnarray}
Recall that all vertices which have the same or greater number of
derivatives as free Lagrangian are always defined up to possible field
redefinitions. For the case at hands we have the following
possibilities:
\begin{equation}
\Omega_\mu{}^{ab} \Rightarrow \Omega_\mu{}^{ab} + \kappa_1
h_\mu{}^{(a} B^{b)} - Tr, \qquad \omega_\mu{}^a \Rightarrow
\omega_\mu{}^a + \kappa_2 \Phi_\mu{}^{ab} B^b
\end{equation}
Let us consider variations under gravitational $\tilde{\xi}^a$
transformations:
\begin{eqnarray*}
\delta_0 {\cal L}_1 &=& \eptwo [ a_1 D_\mu \Omega_\nu{}^{ac} B^b
\tilde{\xi}^c + a_2 D_\mu \Omega_\nu{}^{ac} B^c \tilde{\xi}^b - a_1
\Omega_\mu{}^{ac} \tilde{\xi}^c D_\nu B^b - a_2 \Omega_\mu{}^{ac}
\tilde{\xi}^b D_\nu B^b ] - \\
 && - a_5 \varepsilon^{\mu\nu\alpha} D_\mu \Phi_\nu{}^{ab}
\tilde{\xi}^a D_\alpha B^b
\end{eqnarray*}
To compensate these variations we introduce the following corrections
to gauge transformations:
\begin{eqnarray*}
\delta_1 \Phi_\mu{}^{ab} &=& \alpha_1 \varepsilon_\mu{}^{c(a} B^{b)}
\tilde{\xi}^c + \alpha_2 \varepsilon_\mu{}^{c(a} \tilde{\xi}^{b)} B^c
- Tr \\
\delta_1 \Omega_\mu{}^{ab} &=& \alpha_3 \tilde{\xi}^{(a} D_\mu B^{b)}
- Tr
\end{eqnarray*}
They produce the following variations:
\begin{eqnarray*}
\delta_1 {\cal L}_0 &=& 2\eptwo [ D_\mu \Omega_\nu{}^{ac} ( \alpha_1
B^c \tilde{\xi}^b + \alpha_2 B^b \tilde{\xi}^c) - \alpha_3
\Omega_\mu{}^{ac} (\tilde{\xi}^b D_\nu B^c + \tilde{\xi}^c D_\nu B^b)
] + \\
 && + 2\alpha_3 \varepsilon^{\mu\nu\alpha} D_\mu \Phi_\nu{}^{ab}
\tilde{\xi}^a D_\alpha B^b
\end{eqnarray*}
Thus we obtain
$$
a_1 = a_2 = - 2\alpha_3, \qquad a_5 = 2\alpha_3 \quad \Longrightarrow
\quad a_1 = a_2 = - a_5
$$
but such relations on $a_{1,2,5}$ means that these terms can be
removed by field redefinition with parameter $\kappa_1$. This leaves
us with
$$
{\cal L}_1 = \eptwo [ a_3 \Phi_\mu{}^{ac} \omega_\nu{}^c B^b + a_4
\Phi_\mu{}^{ac} \omega_\nu{}^b B^c ] + a_6 \varepsilon^{\mu\nu\alpha}
\partial_\mu h_\nu{}^a \Phi_\alpha{}^{ab} B^b
$$
Now let us consider variations under $\xi^{ab}$ transformations
\begin{eqnarray*}
\delta_0 {\cal L}_1 &=& \xi^{ac} \eptwo [ - a_3 D_\mu \omega_\nu{}^c
B^b - a_4 D_\mu \omega_\nu{}^b B^c + a_3 \omega_\mu{}^c D_\nu B^b +
a_4 \omega_\mu{}^b D_\nu B^c ] - \\
 && - \varepsilon^{\mu\nu\alpha} D_\mu h_\nu{}^a \xi^{ab} D_\alpha B^b
\end{eqnarray*}
Then we introduce corresponding corrections to gauge transformations
$$
\delta \omega_\mu{}^a = \beta_1 \xi^{ab} \partial_\mu B^b, \qquad
\delta h_\mu{}^a = \beta_2 \varepsilon_\mu{}^{ab} \xi^{bc} B^c
$$
which produce
$$
\delta_1 {\cal L}_0 = \beta_1 \eptwo \omega_\mu{}^a \xi^{bc}
\partial_\nu B^c - \beta_1 \varepsilon^{\mu\nu\alpha} \partial_\nu
h_\nu{}^a \xi^{ab} \partial_\alpha B^b - \beta_2
\varepsilon^{\mu\nu\alpha} D_\mu \omega_\nu{}^a
\varepsilon_\alpha{}^{ab} \xi^{bc} B^c
$$
Thus we obtain
$$
a_3 = 0, \qquad a_4 = \beta_1 = - a_6
$$
and it means that all the remaining terms can be removed by field
redefinition with the parameter $\kappa_2$.

\subsection{Canonical vertices}

In this subsection we consider canonical vertices, i.e. those that
correspond to standard minimal interaction with the replacement of
background frame $e_\mu{}^a$ by dynamical one $e_\mu{}^a-h_\mu{}^a$
and $AdS$ covariant derivatives $D_\mu$ by total Lorentz covariant
one $D_\mu - \omega_\mu$. To be sure that the results obtained are
the most general ones initially we take all such terms with
arbitrary coefficients. Thus the ansatz for cubic Lagrangian will
look as follows:
\begin{eqnarray}
{\cal L}_1 &=& \epthree [c_1 h_\mu{}^a \Omega_\nu{}^{bd}
\Omega_\alpha{}^{cd} + c_2 \omega_\mu{}^a \Omega_\nu{}^{bd}
\Phi_\alpha{}^{cd} + c_3 h_\mu{}^a \Omega_\nu{}^b \Omega_\alpha{}^c +
c_4 f_\mu{}^a \omega_\nu{}^b \Omega_\alpha{}^c] + \nonumber \\
&& + c_5 h B^a B^a + c_6 \varepsilon^{\mu\nu\alpha} h_\mu{}^a B^a
D_\nu A_\alpha + c_7 h \pi^a \pi^a + c_8 \eptwo h_\mu{}^a \pi^b D_\nu
\varphi + \nonumber \\
 && + \varepsilon^{\mu\nu\alpha}[ d_1 h_\mu{}^a \Omega_{\nu}{}^{ab}
f_\alpha{}^b + d_2 h_\mu{}^a \Phi_{\nu}{}^{ab} \Omega_\alpha{}^b + d_3
h_\mu{}^a \Omega_{\nu}{}^a A_\alpha + \nonumber \\
&& \qquad + d_4 h_\mu{}^a f_{\nu}{}^a B_\alpha + d_5 h_\mu{}^a
f_{\nu,\alpha} B^a] + d_6 \eptwo h_\mu{}^a \pi^b A_\nu + \nonumber \\
 && + \epthree [e_1 h_\mu{}^a \Phi_\nu{}^{bd} \Phi_\alpha{}^{cd} + e_2
h_\mu{}^a f_\nu{}^{b} f_\alpha{}^{c}] + e_3 \eptwo h_\mu{}^a f_\nu{}^b
\varphi + e_4 h \varphi^2
\end{eqnarray}
Let us begin with gravitational Lorentz transformations. Calculating
variations under the $\tilde{\eta}^a$ transformations we obtain
\begin{eqnarray*}
\delta_0 {\cal L}_1 &=& \epthree [ - c_2 D_\mu \Omega_\nu{}^{ad}
\Phi_\alpha{}^{bd} \tilde{\eta}^c - c_2 D_\mu \Phi_\nu{}^{ad}
\Omega_\alpha{}^{bd} \tilde{\eta}^c - c_4 D_\mu \Omega_\nu{}^a
f_\alpha{}^b \tilde{\eta}^c - c_4 D_\mu f_\nu{}^a \Omega_\alpha{}^b
\tilde{\eta}^c ] + \\
 && + \varepsilon^{\mu\nu\alpha} [ 2c_1 \Omega_{\mu,\nu}{}^a
\Omega_\alpha{}^{ab} \tilde{\eta}^b + 2c_3 \Omega_{\mu,\nu}
\Omega_\alpha{}^a \tilde{\eta}^a + c_8 \pi_\mu \tilde{\eta}_\nu
D_\alpha \varphi ] + c_6 \eptwo B^a \tilde{\eta}^b D_\mu A_\nu + \\
 && + \eptwo [ d_1 \Omega_\mu{}^{ac} f_\nu{}^c \tilde{\eta}^b + d_2
\Phi_\mu{}^{ac} \Omega_\nu{}^c \tilde{\eta}^b
 + d_3 \Omega_\mu{}^a
A_\nu \tilde{\eta}^b + d_4 f_\mu{}^a B_\nu \tilde{\eta}^b + d_5
f_{\mu,\nu} B^a \tilde{\eta}^b ] + \\
 && + \varepsilon^{\mu\nu\alpha} [ d_6 \pi_\mu \tilde{\eta}_\nu
A_\alpha + 2e_1 \Phi_{\mu,\nu}{}^a \Phi_\alpha{}^{ab} \tilde{\eta}^b +
2e_2 f_{\mu,\nu} f_\alpha{}^a \tilde{\eta}^a + e_3 f_{\mu,\nu}
\tilde{\eta}_\alpha \varphi ]
\end{eqnarray*}
By analyzing the terms with explicit derivatives it is not hard to
determine the corresponding corrections to gauge transformations:
\begin{eqnarray*}
\delta_1 \Omega_\mu{}^{ab} &=& \frac{c_2}{2} \varepsilon^{cd(a}
\Omega_\mu{}^{b)c} \tilde{\eta}^d, \qquad \delta_1 \Phi_\mu{}^{ab} =
\frac{c_2}{2} \varepsilon^{cd(a} \Phi_\mu{}^{b)c} \tilde{\eta}^d \\
\delta_1 \Omega_\mu{}^a &=& - c_4 \varepsilon^{abc} \Omega_\mu{}^b
\tilde{\eta}^c, \qquad \delta_1 f_\mu{}^a = - c_4 \varepsilon^{abc}
f_\mu{}^b \tilde{\eta}^c \\
\delta_1 B^a &=& c_6 \varepsilon^{abc} B^b \tilde{\eta}^c, \qquad
\delta_1 \pi^a = - c_8 \varepsilon^{abc} \pi^b \tilde{\eta}^c
\end{eqnarray*}
They produce
\begin{eqnarray*}
\delta_1 {\cal L}_0 &=& \epthree [ c_2 D_\mu \Omega_\nu{}^{ad}
\Phi_\alpha{}^{bd} \tilde{\eta}^c + c_2 D_\mu \Phi_\nu{}^{ad}
\Omega_\alpha{}^{bd} \tilde{\eta}^c + c_4 D_\mu \Omega_\nu{}^a
f_\alpha{}^b \tilde{\eta}^c + c_4 D_\mu f_\nu{}^a \Omega_\alpha{}^b
\tilde{\eta}^c ] + \\
 && + \varepsilon^{\mu\nu\alpha} [ - c_2 \Omega_{\mu,\nu}{}^a
\Omega_\alpha{}^{ab} \tilde{\eta}^b - c_4 \Omega_{\mu,\nu}
\Omega_\alpha{}^a \tilde{\eta}^a - c_8 \pi_\mu \tilde{\eta}_\nu
D_\alpha \varphi ] - c_6 \eptwo B^a \tilde{\eta}^b D_\mu A_\nu \\
 && - \frac{3mc_2}{2} \eptwo \Omega_\mu{}^{ac} \tilde{\eta}^b
f_\nu{}^c - 3m(\frac{c_2}{2} + c_4) \epthree \Omega_{\mu,\nu}{}^a
f_\alpha{}^b \tilde{\eta}^c - \\
 && - \frac{mc_2}{2} \eptwo \Phi_\mu{}^{ac} \tilde{\eta}^b
\Omega_\nu{}^c - m (\frac{c_2}{2} + c_4) \epthree \Phi_{\mu,\nu}{}^a
\Omega_\alpha{}^b \tilde{\eta}^c - \\
 && - 2\tilde{m}c_4 \eptwo \Omega_\mu{}^a \tilde{\eta}^b A_\nu +
\tilde{m}c_4 \eptwo f_\mu{}^a \tilde{\eta}^b B_\nu + \tilde{m}c_6
\eptwo f_{\mu,\nu} B^a \tilde{\eta}^b - Mc_8
\varepsilon^{\mu\nu\alpha} \pi_\mu \tilde{\eta}_\nu A_\alpha - \\
 && - \varepsilon^{\mu\nu\alpha} [ \frac{M^2c_2}{36}
\Phi_{\mu,\nu}{}^a \Phi_\alpha{}^{ab} \tilde{\eta}^b +
\frac{M^2c_4}{4} f_{\mu,\nu} f_\alpha{}^a \tilde{\eta}^a +
M\tilde{m}c_4 f_{\mu,\nu} \tilde{\eta}_\alpha \varphi ]
\end{eqnarray*}
Then requiring that $\delta_0 {\cal L}_1 + \delta_1 {\cal L}_0 = 0$ we
obtain
$$
c_2 = 2c_1, \qquad 2c_3 = c_4 = - c_1
$$
$$
d_1 = 3mc_1, \quad d_2 = mc_1, \quad d_3 = - 2\tilde{m}c_1, \quad
d_4 = \tilde{m}c_1, \quad d_5 = - \tilde{m}c_6, \quad d_6 = Mc_8
$$
$$
e_1 = \frac{M^2c_1}{36}, \qquad e_2 = - \frac{M^2c_1}{8}, \qquad
e_3 = - M\tilde{m}c_1
$$
Now let us consider variations under the $\tilde{\xi}^a$
transformations
\begin{eqnarray*}
\delta_0 {\cal L}_1 &=& \epthree [ - 2c_1 D_\mu \Omega_\nu{}^{ad}
\Omega_\alpha{}^{bd} \tilde{\xi}^c - 2c_3 D_\mu \Omega_\nu{}^a
\Omega_\alpha{}^b \tilde{\xi}^c ] - \\
 && - 2c_5 \tilde{\xi}^\mu B^a D_\mu B^a - c_6
\varepsilon^{\mu\nu\alpha} \tilde{\xi}^a D_\mu B^a D_\nu A_\alpha -
2c_7 \tilde{\xi}^\mu \pi^a D_\mu \pi^a - c_8 \eptwo \tilde{\xi}^a
D_\mu \pi^b D_\nu \varphi + \\
 && + \varepsilon^{\mu\nu\alpha} [ - d_1 \tilde{\xi}^a D_\mu
\Omega_\nu{}^{ab} f_\alpha{}^b + d_1 \tilde{\xi}^a \Omega_\mu{}^{ab}
D_\nu f_\alpha{}^b - d_2 \tilde{\xi}^a D_\mu \Phi_\nu{}^{ab}
\Omega_\alpha{}^b + d_2 \tilde{\xi}^a \Phi_\mu{}^{ab} D_\nu
\Omega_\alpha{}^b - \\
 && - d_3 \tilde{\xi}^a D_\mu \Omega_\nu{}^a A_\alpha + d_3
\tilde{\xi}^a \Omega_\mu{}^a D_\nu A_\alpha - d_4 \tilde{\xi}^a D_\mu
f_\nu{}^a B_\alpha + d_4 \tilde{\xi}^a f_\mu{}^a D_\nu B_\alpha - \\
 && - d_5 \tilde{\xi}^a D_\mu f_{\nu,\alpha} B^a - d_5 \tilde{\xi}^a
f_{\mu,\nu} D_\alpha B^a ] - d_6 \eptwo [ \tilde{\xi}^a D_\mu \pi^b
A_\nu + \tilde{\xi}^a \pi^b D_\mu A_\nu ] - \\
 && - \epthree [ 2e_1 D_\mu \Phi_\nu{}^{ad} \Phi_\alpha{}^{bd}
\tilde{\xi}^c + 2e_2 D_\mu f_\nu{}^a f_\alpha{}^b
\tilde{\xi}^c ] + e_3 \eptwo [ - \tilde{\xi}^a D_\mu f_\nu{}^a \varphi
+ \tilde{\xi}^a f_\mu{}^b D_\nu \varphi ] - \\
 && - 2e_4 \tilde{\xi}^\mu \varphi D_\mu \varphi  + \lambda^2
\varepsilon^{\mu\nu\alpha} [ c_2 (\Omega_{\mu,\nu}{}^a
\Phi_\alpha{}^{ab} + \Omega_\mu{}^{ab} \Phi_{\nu,\alpha}{}^a )
\tilde{\xi}^b + c_4 (\Omega_{\mu,\nu} f_\alpha{}^a + \Omega_\mu{}^a
f_{\nu,\alpha}) \tilde{\xi}^a ]
\end{eqnarray*}
In this case we have a lot of terms with explicit derivatives and the
corresponding corrections to gauge transformations look as follows:
\begin{eqnarray*}
\delta \Omega_\mu{}^{ab} &=& \frac{d_2}{2} (\Omega_\mu{}^{(a}
\tilde{\xi}^{b)} - \frac{2}{3} g^{ab} \Omega_\mu{}^c \tilde{\xi}^c) +
e_1 \varepsilon^{cd(a} \Phi_\mu{}^{b)c} \tilde{\xi}^d \\
\delta \Phi_\mu{}^{ab} &=& c_1 \varepsilon^{cd(a} \Omega_\mu{}^{b)c}
\tilde{\xi}^d + \frac{d_1}{2} (f_\mu{}^{(a}
\tilde{\xi}^{b)} - \frac{2}{3} g^{ab} f_\mu{}^c \tilde{\xi}^c) \\
\delta \Omega_\mu{}^a &=& d_1 \Omega_\mu{}^{ab} \tilde{\xi}^b - d_4
B_\mu \tilde{\xi}^a - d_5 e_\mu{}^a B^b \tilde{\xi}^b - 2e_2
\varepsilon^{abc} f_\mu{}^b \tilde{\xi}^c \\
\delta f_\mu{}^a &=& c_1 \varepsilon^{abc} \Omega_\mu{}^b
\tilde{\xi}^c + d_2 \Phi_\mu{}^{ab} \tilde{\xi}^b - d_3 A_\mu
\tilde{\xi}^a \\
\delta B_\mu &=& - c_6 \tilde{\xi}^a D_\mu B^a + d_3 \Omega_\mu{}^a
\tilde{\xi}^a + d_6 \varepsilon_\mu{}^{ab} \pi^a \tilde{\xi}^b \\
\delta A_\mu &=& - c_6 \varepsilon_\mu{}^{ab} B^a \tilde{\xi}^b +
d_4 f_\mu{}^a \tilde{\xi}^a \\
\delta \pi^a &=& c_8 ( \tilde{\xi}^\mu D_\mu \pi^a - \tilde{\xi}^a
(D\pi)) + e_3 (f \tilde{\xi}^a - f_\mu{}^a \tilde{\xi}^\mu) +
2e_4 \varphi \tilde{\xi}^a \\
\delta \varphi &=& c_8 (\pi\tilde{\xi})
\end{eqnarray*}
By straightforward but rather long calculations we can see that all
variations can be canceled provided the following relations hold:
$$
- 2c_5 = c_6 = 2c_7 = - c_8 = c_1, \qquad e_4 = - 3\tilde{m}^2 c_1
$$
Thus the invariance under the gravitational gauge transformations
already completely fixes all the coefficients in the Lagrangian and in
the corresponding corrections to gauge transformations up to one
arbitrary coupling constant. Choosing the same coupling constant
$\alpha_1$ as in the massless spin 3 case we obtain the following
final form of the cubic vertex (compare with the free Lagrangian
(\ref{s3l0})):
\begin{eqnarray}
{\cal L}_1 &=& \alpha_1 \epthree [ h_\mu{}^a \Omega_\nu{}^{bd}
\Omega_\alpha{}^{cd} + 2 \omega_\mu{}^a \Omega_\nu{}^{bd}
\Phi_\alpha{}^{cd} - \frac{1}{2} h_\mu{}^a \Omega_\nu{}^b
\Omega_\alpha{}^c - f_\mu{}^a \omega_\nu{}^b \Omega_\alpha{}^c] +
\nonumber \\
&& + \alpha_1 [ - \frac{1}{2} h B^a B^a + \varepsilon^{\mu\nu\alpha}
h_\mu{}^a B^a D_\nu A_\alpha + \frac{1}{2} h \pi^a \pi^a - \eptwo
h_\mu{}^a \pi^b D_\nu \varphi ] + \nonumber \\
 && + \alpha_1 \varepsilon^{\mu\nu\alpha}[ 3m h_\mu{}^a
\Omega_{\nu}{}^{ab} f_\alpha{}^b + m h_\mu{}^a \Phi_{\nu}{}^{ab}
\Omega_\alpha{}^b + 2\tilde{m} h_\mu{}^a \Omega_{\nu}{}^a A_\alpha -
\nonumber \\
&& \qquad \qquad - \tilde{m} h_\mu{}^a f_{\nu}{}^a B_\alpha -
\tilde{m} h_\mu{}^a f_{\nu,\alpha} B^a] - M\alpha_1 \eptwo h_\mu{}^a
\pi^b A_\nu + \nonumber \\
 && + \alpha_1 \epthree [ \frac{M^2}{36} h_\mu{}^a \Phi_\nu{}^{bd}
\Phi_\alpha{}^{cd} - \frac{M^2}{8} h_\mu{}^a f_\nu{}^{b}
f_\alpha{}^{c}] - \nonumber \\
 && - M\tilde{m}\alpha_1 \eptwo h_\mu{}^a f_\nu{}^b \varphi -
3\tilde{m}^2\alpha_1 h \varphi^2
\end{eqnarray}
and corrections to the gravitational gauge transformations:
\begin{eqnarray}
\delta_1 \Omega_\mu{}^{ab} &=& \alpha_1 [ \varepsilon^{cd(a}
\Omega_\mu{}^{b)c} \tilde{\eta}^d + \frac{m}{2} (\Omega_\mu{}^{(a}
\tilde{\xi}^{b)} - \frac{2}{3} g^{ab} \Omega_\mu{}^c \tilde{\xi}^c) +
\frac{M^2}{36} \varepsilon^{cd(a} \Phi_\mu{}^{b)c} \tilde{\xi}^d ]
\nonumber \\
\delta_1 \Phi_\mu{}^{ab} &=& \alpha_1 [ \varepsilon^{cd(a}
\Phi_\mu{}^{b)c} \tilde{\eta}^d + \varepsilon^{cd(a}
\Omega_\mu{}^{b)c} \tilde{\xi}^d + \frac{3m}{2} (f_\mu{}^{(a}
\tilde{\xi}^{b)} - \frac{2}{3} g^{ab} f_\mu{}^c \tilde{\xi}^c) ]
\nonumber \\
\delta_1 \Omega_\mu{}^a &=& \alpha_1 [ \varepsilon^{abc}
\Omega_\mu{}^b
\tilde{\eta}^c + 3m \Omega_\mu{}^{ab} \tilde{\xi}^b - \tilde{m}
B_\mu \tilde{\xi}^a + \tilde{m} e_\mu{}^a B^b \tilde{\xi}^b +
\frac{M^2}{4} \varepsilon^{abc} f_\mu{}^b \tilde{\xi}^c - M\tilde{m}
\varepsilon_\mu{}^{ab} \tilde{\xi}^b \varphi ] \nonumber \\
\delta_1 f_\mu{}^a &=& \alpha_1 [ \varepsilon^{abc} f_\mu{}^b
\tilde{\eta}^c + \varepsilon^{abc} \Omega_\mu{}^b \tilde{\xi}^c + m
\Phi_\mu{}^{ab} \tilde{\xi}^b + 2\tilde{m} A_\mu \tilde{\xi}^a ]
\nonumber \\
\delta_1 B_\mu &=& \alpha_1 [ \varepsilon_\mu{}^{ab} B^a
\tilde{\eta}^b - \tilde{\xi}^a D_\mu B^a - 2\tilde{m} \Omega_\mu{}^a
\tilde{\xi}^a - M \varepsilon_\mu{}^{ab} \pi^a \tilde{\xi}^b ]
\label{g1} \\
\delta_1 A_\mu &=& \alpha_1 [ - \varepsilon_\mu{}^{ab} B^a
\tilde{\xi}^b + \tilde{m} f_\mu{}^a \tilde{\xi}^a ] \nonumber \\
\delta_1 \pi^a &=& \alpha_1 [ \varepsilon^{abc} \pi^b \tilde{\eta}^c -
( \tilde{\xi}^\mu D_\mu \pi^a - \tilde{\xi}^a (D\pi)) - M\tilde{m}
(f \tilde{\xi}^a - f_\mu{}^a \tilde{\xi}^\mu) - 6\tilde{m}^2 \varphi
\tilde{\xi}^a ] \nonumber \\
\delta_1 \varphi &=& - \alpha_1 (\pi\tilde{\xi}) \nonumber
\end{eqnarray}
But we still have to take care on the whole set of massive spin 3
gauge transformations $\eta^{ab}$, $\xi^{ab}$, $\eta^a$, $\xi^a$,
$\xi$. Note that there is a striking difference between $d=3$ and $d
\ge 4$ cases here. In $d \ge 4$ dimensions variations of the
Lagrangian under the higher spin gauge transformations produce terms
proportional to gravitational curvature and to compensate for these
terms one has to introduce a lot of higher derivative corrections to
the Lagrangian and gauge transformations \cite{FV87,FV87a}.
Moreover, the higher the spin one will try to consider the more
derivatives one will have to introduce. In this, the most important
terms are the ones containing Weyl tensor because all the terms
containing Ricci tensor can be compensated by corresponding
corrections to graviton transformations and/or removed by field
redefinitions. The same is true for the massive higher spins as well.
Complete analysis for massive spin-3 fields gravitational interactions
in $d \ge 4$ dimensions is still absent but the results of
\cite{Zin08} clearly shows that in this case the four derivatives
terms containing full Riemann tensor play crucial role. But in three
dimensions Weyl tensor is identically zero and as a result, as we will
show on the example of massive spin 3 field now, it is possible to
achieve gauge invariance in the linear approximation without
introduction of any higher derivative corrections. Thus after
switching on gravitational interactions we still can consider any
massless, partially massless or flat limits. Moreover, as it will be
seen from the next section, this holds true for the arbitrary spins as
well.

So our cubic vertex is completely fixed and now we have to find
corrections to massive spin-3 gauge transformations to achieve
invariance in the linear approximation. The procedure goes exactly
in the same way as it was shown above: we calculate variations of
cubic vertex $\delta_0 {\cal L}_1$, analyze the terms containing
explicit derivatives, find the form of the necessary corrections to
gauge transformations and adjust coefficients to achieve $\delta_0
{\cal L}_1 + \delta_1 {\cal L}_0 = 0$. All the calculations are
straightforward but rather long so we will not give the details of
them here and simply give the final results. For massive spin 3
fields the corrections to gauge transformations appear to be:
\begin{eqnarray}
\delta_1 \Omega_\mu{}^{ab} &=& \alpha_1 [ - \varepsilon^{cd(a}
\eta^{b)c} \omega_\mu{}^d - \frac{m}{2} (h_\mu{}^{(a} \eta^{b)} -
\frac{2}{3} g^{ab} h_\mu{}^c \eta^c) - \frac{M^2}{36}
\varepsilon^{cd(a} \xi^{b)c} h_\mu{}^d ] \nonumber \\
\delta_1 \Phi_\mu{}^{ab} &=& \alpha_1 [ - \varepsilon^{cd(a}
\eta^{b)c} h_\mu{}^d - \varepsilon^{cd(a} \xi^{b)c} \omega_\mu{}^d
- \frac{3m}{2} (h_\mu{}^{(a} \xi^{b)} - \frac{2}{3} g^{ab} h_\mu{}^c
\xi^c) ] \nonumber \\
\delta_1 \Omega_\mu{}^a &=& \alpha_1 [ \varepsilon^{abc}
\omega_\mu{}^b \eta^c - 3m \eta^{ab} h_\mu{}^b + \frac{M^2}{4}
\varepsilon^{abc} h_\mu{}^b \xi^c ] \label{g2} \\
\delta_1 f_\mu{}^a &=& \alpha_1 [ \varepsilon^{abc} h_\mu{}^b \eta^c +
\varepsilon^{abc} \omega_\mu{}^b \xi^c - m \xi^{ab} h_\mu{}^b -
2\tilde{m} h_\mu{}^a \xi ] \nonumber \\
\delta_1 A_\mu &=& - \alpha_1 \tilde{m} h_\mu{}^a \xi^a \nonumber
\end{eqnarray}
Comparing these expressions with the initial gauge transformations
(\ref{s3d0}), one can see that they indeed correspond to standard
substitution rules. At the same time the gravitational fields also
have non-trivial transformations:
\begin{eqnarray}
\delta_1 \omega_\mu{}^a &=& \alpha_1 [ - 2 \varepsilon^{abc}
\Omega_\mu{}^{bd} \eta^{cd} + \varepsilon^{abc} \Omega_\mu{}^b \eta^c
+ 3m \eta^{ab} f_\mu{}^b + m \xi^{ab} \Omega_\mu{}^b - m
\Phi_\mu{}^{ab} \eta^b - 2\tilde{m} A_\mu \eta^a - \nonumber \\
 && - 3m \Omega_\mu{}^{ab} \xi^b + \tilde{m} B_\mu \xi^a +
2\tilde{m} \Omega_\mu{}^a \xi - \frac{M^2}{18} \varepsilon^{abc}
\Phi_\mu{}^{bd} \xi^{cd} + \frac{M^2}{4} \varepsilon^{abc} f_\mu{}^b
\xi^c - M\tilde{m} \varepsilon_\mu{}^{ab} \xi^b \varphi ]  \nonumber
\\
\delta_1 h_\mu{}^a &=& \alpha_1 \varepsilon^{abc} [ - 2
\Phi_\mu{}^{bd} \eta^{cd} - 2 \Omega_\mu{}^{bd} \xi^{cd} + f_\mu{}^b
\eta^c + \Omega_\mu{}^b \xi^c ] \label{g3}
\end{eqnarray}

Having in our disposal linear corrections to all gauge transformations
we can calculate their commutators in the lowest order and extract
that composition laws for gauge parameters showing us the structure
of gauge algebra behind these transformations. For the massive spin-3
gauge parameters form (\ref{g1}) and (\ref{g2}) we obtain (up to the
common multiplier $\alpha_1{}^2$):
\begin{eqnarray}
\hat{\eta}^{ab} &=& \varepsilon^{cd(a}  ( \eta^{b)c} \tilde{\eta}^c +
\frac{M^2}{36} \xi^{b)c} \tilde{\xi}^c ) + \frac{m}{2} ( \xi^{(a}
\tilde{\xi}^{b)} - Tr) \nonumber \\
\hat{\xi}^{ab} &=& \varepsilon^{cd(a} ( \xi^{b)c} \tilde{\eta}^c +
\eta^{b)c} \tilde{\xi}^c) + \frac{3m}{2} ( \xi^{(a} \tilde{\xi}^{b)} -
Tr) \nonumber \\ 
\hat{\eta}^a &=& \varepsilon^{abc} ( \eta^b \tilde{\eta}^c +
\frac{M^2}{4} \xi^b \tilde{\xi}^c) + 3m \eta^{ab} \tilde{\xi}^b \\
\hat{\xi}^a &=& \varepsilon^{abc} ( \xi^b \tilde{\eta}^c + \eta^b
\tilde{\xi}^c) + m \xi^{ab} \tilde{\xi}^b + 2\tilde{m} \xi
\tilde{\xi}^a \nonumber \\
\hat{\xi} &=& \tilde{m} \xi^a \tilde{\xi}^a \nonumber
\end{eqnarray}
and for the graviton gauge parameters, correspondingly:
\begin{eqnarray}
\hat{\tilde{\eta}}^a &=& \varepsilon^{abc}
( -2 \eta_1{}^{bd} \eta_2{}^{cd} - \frac{M^2}{18} \xi_1{}^{bd}
\xi_2{}^{cd} + \eta_1{}^b \eta_2{}^c + \frac{M^2}{4} \xi_1{}^b
\xi_2{}^c ) + \nonumber \\
 && \qquad + 3m \eta^{ab} \xi^b + m \xi^{ab} \eta^b + 2\tilde{m}
\eta^a \xi \\
\hat{\tilde{\xi}}^a &=& \varepsilon^{abc} ( - 2 \eta^{bd} \xi^{cd} +
\eta^b \xi^c ) \nonumber
\end{eqnarray}
We see that the structure of the algebra drastically differs from that
of the pure massless spin-3 case interacting with gravity. The main
reason is that Stueckelberg spin-2 and spin-1 fields are gauge fields
themselves having their own gauge transformations so that the algebra
must necessarily be extended with corresponding generators. We have
already mentioned that in the massless limit massive spin-3 particles
decomposes into massless spin-3 and massive spin-2 ones. Due to
universality of gravitational interactions even in the massless limit
$m = 0$ (in this $M^2 = \lambda^2$) this massive spin-2 field is still
present and we obtain different algebra.

As in the massless case the invariance of the Lagrangian implies
that there exist deformations for all gauge invariant curvatures
such that under corrected gauge transformations they transform
covariantly and this provides highly non-trivial check for all our
calculations. All additional terms for the massive spin-3 curvatures
contain one forms $\omega_\mu{}^a$ or $h_\mu{}^a$ and completely
determined by the structure of gauge transformations. The results
\begin{eqnarray}
\Delta {\cal G}_{\mu\nu}{}^{ab} &=& \alpha_1 [ \varepsilon^{cd(a}
\Omega_{[\mu}{}^{b)c} \omega_{\nu]}{}^d + \frac{m}{2} (
\Omega_{[\mu}{}^{(a} h_{\nu]}{}^{b)} - \frac{2}{3} g^{ab}
\Omega_{[\mu}{}^c h_{\nu]}{}^c) + \frac{M^2}{36} \varepsilon^{cd(a}
\Phi_{[\mu}{}^{b)c} h_{\nu]}{}^d ] \nonumber \\
\Delta {\cal H}_{\mu\nu}{}^{ab} &=& \alpha_1 [ \varepsilon^{cd(a}
(\Phi_{[\mu}{}^{b)c} \omega_{\nu]}{}^d + \Omega_{[\mu}{}^{b)c}
h_{\nu]}{}^d) + \frac{3m}{2} (f_{[\mu}{}^{(a} h_{\nu]}{}^{b)} -
\frac{2}{3} g^{ab} f_{[\mu}{}^c h_{\nu]}{}^c) ] \nonumber \\
\Delta {\cal F}_{\mu\nu}{}^a &=& \alpha_1 [ \varepsilon^{abc}
\Omega_{[\mu}{}^b \omega_{\nu]}{}^c + 3m \Omega_{[\mu}{}^{ab}
h_{\nu]}{}^b - \tilde{m} B_{[\mu} h_{\nu]}{}^a + \tilde{m}
e_{[\mu}{}^a h_{\nu]}{}^b B^b + \nonumber \\
 && \quad + \frac{M^2}{4} \varepsilon^{abc} f_{[\mu}{}^b h_{\nu]}{}^c
- M\tilde{m} \varepsilon_{[\mu}{}^{ab} h_{\nu]}{}^b \varphi ] \\
\Delta {\cal T}_{\mu\nu}{}^a &=& \alpha_1 [ \varepsilon^{abc}
(f_{[\mu}{}^b \omega_{\nu]}{}^c + \Omega_{[\mu}{}^b h_{\nu]}{}^c) + m
\Phi_{[\mu}{}^{ab} h_{\nu]}{}^b + 2\tilde{m} A_{[\mu} h_{\nu]}{}^a ]
\nonumber \\
\Delta {\cal B}_\mu{}^a &=& \alpha_1 [ \varepsilon^{abc}
\omega_\mu{}^b B^c + \frac{M}{2} \varepsilon^{abc} \pi^b h_\mu{}^c +
V^{ab} h_\mu{}^b ] \nonumber \\
\Delta {\cal A}_{\mu\nu} &=& \alpha_1 [ - \varepsilon_{[\mu}{}^{ab}
B^a h_{\nu]}{}^b + \tilde{m} f_{[\mu}{}^a h_{\nu]}{}^a ] \nonumber \\
\Delta \Pi_\mu{}^a &=& \alpha_1 [ \varepsilon^{abc} \omega_\mu{}^b
\pi^c + 2\tilde{m}^2 \varphi h_\mu{}^a + W^{ab} h_\mu{}^b ], \qquad
\Delta \Phi_\mu = \alpha_1 h_\mu{}^a \pi^a \nonumber
\end{eqnarray}
clearly show that these deformations (as well as corrections to gauge
transformations) exactly correspond to standard  substitution rules.
As for the deformations for gravitational curvature and torsion, all
terms containing one forms are also determined by the very structure
of corrections to gauge transformations, while the terms quadratic in
zero forms are determined by the requirement that deformed curvatures
transform covariantly. By straightforward but rather lengthy
calculations we obtain
\begin{eqnarray}
\Delta R_{\mu\nu}{}^a &=& \alpha_1 [ - \varepsilon^{abc}
\Omega_{[\mu}{}^{bd} \Omega_{\nu]}{}^{cd} + \frac{1}{2}
\Omega_{[\mu}{}^b \Omega_{\nu]}{}^c - 3m \Omega_{[\mu}{}^{ab}
f_{\nu]}{}^b - m \Phi_{[\mu}{}^{ab} \Omega_{\nu]}{}^b + 2\tilde{m}
\Omega_{[\mu}{}^a A_{\nu]} - \nonumber \\
 && \quad - \tilde{m} f_{[\mu}{}^a B_{\nu]} - \frac{M^2}{36}
\varepsilon^{abc} \Phi_{[\mu}{}^{bd} \Phi_{\nu]}{}^{cd} +
\frac{M^2}{8} \varepsilon^{abc} f_{[\mu}{}^b f_{\nu]}{}^c - M\tilde{m}
\varepsilon_{[\mu}{}^{ab} f_{\nu]}{}^b \varphi - \nonumber \\
 && \quad - \varepsilon_{\mu\nu}{}^b B^a B^b + \frac{1}{2}
\varepsilon_{\mu\nu}{}^a B^2 - \varepsilon_{\mu\nu}{}^b \pi^a \pi^b +
\frac{1}{2} \varepsilon_{\mu\nu}{}^a \pi^2 + 3\tilde{m}^2
\varepsilon_{\mu\nu}{}^a \varphi^2 ] \\
\Delta T_{\mu\nu}{}^a &=& \alpha_1 \varepsilon^{abc} [ - 2
\Omega_{[\mu}{}^{bd} \Phi_{\nu]}{}^{cd} + \Omega_{[\mu}{}^b
f_{\nu]}{}^c ] \nonumber
\end{eqnarray}

We have already mentioned that the system gravity plus massless spin 3
provides closed consistent theory without any needs to introduce some
other fields. But if we consider quadratic variations to check if
$\delta_1 {\cal L}_1 = 0$ we immediately recover that contrary to the
massless case the system gravity plus massive spin 3 is not closed.
The crucial point is the variations under the main spin 3
transformations $\eta^{ab}$ and $\xi^{ab}$ that contain one forms
$\Omega_\mu{}^{ab}$, $\Phi_\mu{}^{ab}$, $\Omega_\mu{}^a$ and
$f_\mu{}^a$. In three dimensions there are no any quartic vertices
constructed out of one forms, while zero forms $B^a$, $\pi^a$ and
$\varphi$ do not have non-trivial transformations and can not be of
any help. Thus to construct consistent theory (if it exists at all) we
must introduce other fields into the system\footnote{In the case
under consideration, the most natural candidate for the role of the
new field is massive spin-4 field which will generate the new
interaction terms. In turn the new massive spin-4 field almost
certainly will require the introduction of even higher spins and so on
up to infinity. Therefore one can suggest that the nonlinear theory of
massive higher-spin fields does not have a truncated version, unlike
to massless case where we can use a finite set of fields}. Let us
stress however that the existence and the very structure of cubic
vertex do not depend on the presence or absence of any other fields so
that the results obtained here are universal and model independent.

\section{Massive arbitrary spin-$s$ field}

In this section we consider cubic gravitational vertex for massive
field with arbitrary integer spin $s$. We already know that the system
gravity plus massless spin $s$ is not closed so that there is no
chance that it can be the case for massive field. But as we have
already mentioned the existence and structure of cubic vertex do not
depend on the presence of any other fields. Moreover the mere
existence of such cubic vertex is crucial for any further
investigations.

\subsection{Kinematics}

We will use frame-like gauge invariant description for massive spin
$s$ field elaborated in our previous work \cite{BSZ12a}. Such
description uses a collection of one forms $\Omega_\mu{}^{(k)}$,
$\Phi_\mu{}^{(k)}$, $1 \le k \le s-1$ (in the same compact notations)
and a pair of zero forms $B^a$, $\pi^a$. The free Lagrangian has the
form:
\begin{eqnarray}
{\cal L}_0 &=& \sum_{k=1}^{s-1} (-1)^{k+1} [ \frac{k}{2} \eptwo (
\Omega_\mu{}^{a(k-1)} \Omega_\nu{}^{b(k-1)} + \beta_k{}^2
\Phi_\mu{}^{a(k-1)} \Phi_\nu{}^{b(k-1)} ) - \varepsilon^{\mu\nu\alpha}
\Omega_\mu{}^{(k)} D_\nu \Phi_\alpha{}^{(k)} ] - \nonumber \\
 && - \sum_{k=2}^{s-1} (-1)^{k+1} k\alpha_k \varepsilon^{\mu\nu\alpha}
[ \frac{k+1}{k-1} \Omega_{\mu,\nu}{}^{(k-1)} \Phi_\alpha{}^{(k-1)} +
\Omega_\mu{}^{(k-1)} \Phi_{\nu,\alpha}{}^{(k-1)} ] \nonumber + \\
 && + \frac{1}{2} B^a B^a - \varepsilon^{\mu\nu\alpha} [ B_\mu D_\nu
\pi_\alpha + \frac{\gamma_0}{2} B_\mu \Phi_{\nu,\alpha} - \gamma_0
\Omega_{\mu,\nu} \pi_\alpha ] + 2\beta_1{}^2 \pi^a \pi^a
\end{eqnarray}
and is invariant under the following gauge transformations:
\begin{eqnarray}
\delta_0 \Omega_\mu{}^{(k)} &=& D_\mu \eta^{(k)} +
\alpha_k [e_\mu{}^{(1} \eta^{k-1)} -
\frac{2}{2k-1} g^{(2} \eta_\mu{}^{k-2)}]
- \beta_k{}^2 \varepsilon_\mu{}^{b(1}
\xi^{k-1)b} + \gamma_k \eta_\mu{}^{(k)} \nonumber \\
\delta_0 \Phi_\mu{}^{(k)} &=& D_\mu \xi^{(k)} -
\varepsilon_\mu{}^{b(1} \eta^{k-1)b} +
\alpha_k \frac{k+1}{k-1} [e_\mu{}^{(1} \xi^{k-1)} -
\frac{2}{2k-1} g^{(2} \xi_\mu{}^{k-2)}]
 + \frac{k\gamma_k}{k+2} \xi_\mu{}^{(k)} \nonumber \\
\delta_0 B^a &=& \gamma_0 \eta^a, \qquad
\delta_0 \pi^a = - \frac{\gamma_0}{2} \xi^a  \label{arbd0}
\end{eqnarray}
where
\begin{eqnarray*}
\alpha_k{}^2 &=& \frac{(k-1)(s-k)(s+k)}{k^3(k+1)(2k+1)} \left[
\hat{\alpha}_{s-1}{}^2 + (s-k-1)(s+k-1)\lambda^2 \right],\quad
k=2,3...(s-1) \\
\beta_{k} &=& \frac{s(s-1)}{k(k+1)} \beta_{s-1}, \quad
k=1,2...(s-1),\qquad
\beta_{s-1}{}^2 = \frac{s(s-1)}{s-2} \alpha_{s-1}{}^2 + \lambda^2 \\
\gamma_{k} &=& \frac{(k+1)(k+2)}{k} \alpha_{k+1} \quad k=1,2...(s-2)
\\
\gamma_0{}^2 &=& \frac{2(s-1)(s+1)}{3} [\hat{\alpha}_{s-1}{}^2 +
s(s-2)\lambda^2] \\
\hat{\alpha}_{s-1}{}^2 &=& \frac{s(s-1)^3}{s-2} \alpha_{s-1}{}^2
\end{eqnarray*}
Recall that in a de Sitter space ($\Lambda = - \lambda^2 > 0$) there
exists a number of so called partially massless limits
\cite{DW01a,DW01c,Zin01,Zin08b}. It happens each time when one of the
parameters $\alpha_k$ goes to zero. In this, the whole system
decomposes into two independent subsystems, one of them describing
partially massless spin s particle, while the other subsystem
describes massive spin k one.

\subsection{Cubic vertex}

Let us turn to the gravitational interactions for such massive spin
$s$ field. As we have already mentioned, in three dimensions even
for arbitrary spin it is possible to achieve gauge invariance in the
linear approximation without introduction of any higher derivatives
corrections to the Lagrangian and/or gauge transformations. This
time from the very beginning we take cubic vertex corresponding to
standard substitution rules, i.e. the replacement of $AdS$
derivative $D_\mu$ by totally covariant one $D_\mu - \omega_\mu$ and
the background frame $e_\mu{}^a$ by dynamical one $e_\mu{}^a -
h_\mu{}^a$, though we have explicitly checked that it is indeed the
most general solution. Such vertex is determined up to one arbitrary
coupling constant, that for simplicity we set to be 1, and looks as
follows:
\begin{eqnarray}
{\cal L}_1 &=& \sum_{k=1}^{s-1} (-1)^k k \epthree [ \frac{1}{2}
h_\mu{}^a \Omega_\nu{}^{b(k-1)} \Omega_\alpha{}^{c(k-1)} +
\omega_\mu{}^a \Omega_\nu{}^{b(k-1)} \Phi_\alpha{}^{c(k-1)} ] -
\nonumber \\
 && - \frac{1}{2} h B^a B^a + \eptwo \omega_\mu{}^a B_\nu \pi^b +
\varepsilon^{\mu\nu\alpha} h_\mu{}^a [ B^a D_\nu \pi_\alpha + B_\nu
D_\alpha \pi^a ] \nonumber + \\
 && + \sum_{k=2}^{s-1} (-1)^k k\alpha_k \varepsilon^{\mu\nu\alpha} [
\frac{k+1}{k-1} h_\mu{}^a \Omega_\nu{}^{a(k-1)} \Phi_\alpha{}^{(k-1)}
- h_\mu{}^a \Omega_\nu{}^{(k-1)} \Phi_\alpha{}^{a(k-1)} ] + \nonumber
\\
 && + \gamma_0 \varepsilon^{\mu\nu\alpha} h_\mu{}^a [ \Omega_\nu{}^a
\pi_\alpha + \frac{1}{2} B_\nu \Phi_\alpha{}^a -
\Omega_{\nu,\alpha} \pi^a + \frac{1}{2} B^a \Phi_{\nu,\alpha} ] +
\nonumber \\
 && + \sum_{k=1}^{s-1} (-1)^k \frac{k\beta_k{}^2}{2} \epthree
h_\mu{}^a \Phi_\nu{}^{b(k-1)} \Phi_\alpha{}^{c(k-1)} - 2\beta_1{}^2
 h \pi^a \pi^a
\end{eqnarray}
Note that as in the spin-3 case the structure of the vertex is
completely determined (up to possible field redefinitions) by the
invariance under the gravitational $\tilde{\eta}^a$ and
$\tilde{\xi}^a$ transformations. In this, appropriate corrections to
gauge transformations is found to be:
\begin{eqnarray}
\delta_1 \Omega_\mu{}^{(k)} &=& \varepsilon^{ab(1} [
\Omega_\mu{}^{k-1)a} \tilde{\eta}^b + \beta_k{}^2 \Phi_\mu{}^{k-1)a}
\tilde{\xi}^b ] + \frac{(k+1)(k+2)}{k} \alpha_{k+1}
\Omega_\mu{}^{a(k)} \tilde{\xi}^a + \nonumber \\
 && + \alpha_k [ \Omega_\mu{}^{(k-1} \tilde{\xi}^{1)} - \frac{2}{2k-1}
g^{(2} \Omega_\mu{}^{k-2)a} \tilde{\xi}^a ] + \nonumber \\
\delta_1 \Phi_\mu{}^{(k)} &=& \varepsilon^{ab(1} [ \Phi_\mu{}^{k-1)a}
\tilde{\eta}^b + \Omega_\mu{}^{k-1)a} \tilde{\xi}^b
] + (k+1) \alpha_{k+1} \Phi_\mu{}^{a(k)} \tilde{\xi}^a + \nonumber \\
 && + \frac{k+1}{k-1} \alpha_k [ \Phi_\mu{}^{(k-1}
\tilde{\xi}^{1)} - \frac{2}{2k-1} g^{(2} \Phi_\mu{}^{k-2)a}
\tilde{\xi}^a ]  \nonumber \\
\delta_1 \Omega_\mu{}^a &=& \varepsilon^{abc} \Omega_\mu{}^b
\tilde{\eta}^c + 6\alpha_2 \Omega_\mu{}^{ab} \tilde{\xi}^b -
\frac{\gamma_0}{2} e_\mu{}^a B^b \tilde{\xi}^b \\
\delta \Phi_\mu{}^a &=& \varepsilon^{abc} \Phi_\mu{}^b \tilde{\eta}^c
+ 2\alpha_2 \Phi_\mu{}^{ab} \tilde{\xi}^b + \gamma_0e_\mu{}^a \pi^b
\tilde{\xi}^b  \nonumber \\
\delta_1 B_\mu &=& \varepsilon_\mu{}^{ab} B^a \tilde{\eta}^b -
\tilde{\xi}^a D_\mu B^a + \gamma_0 \Omega_\mu{}^a \tilde{\xi}^a - 4
\beta_1{}^2 \varepsilon_\mu{}^{ab} \pi^a \tilde{\xi}^b \nonumber \\
\delta_1 \pi_\mu &=& \varepsilon_\mu{}^{ab} \pi^a \tilde{\eta}^b -
\tilde{\xi}^a D_\mu \pi^a - \varepsilon_\mu{}^{ab} B^a \tilde{\xi}^b -
\frac{\gamma_0}{2} \Phi_\mu{}^a \tilde{\xi}^a \nonumber
\end{eqnarray}
Now we have to take care on all massive spin gauge transformations and
check if we indeed have invariance in the linear approximation. As for
the higher spin fields, their corrections to gauge transformations
again exactly correspond to standard substitution rules:
\begin{eqnarray}
\delta_1 \Omega_\mu{}^{(k)} &=& - \varepsilon^{ab(1} [ \eta^{k-1)a}
\omega_\mu{}^b + \beta_k{}^2 \xi^{k-1)a} h_\mu{}^b ] -
\frac{(k+1)(k+2)}{k} \alpha_{k+1} h_\mu{}^a \eta^{a(k)} - \nonumber \\
 && - k \alpha_k [ h_\mu{}^{(1} \eta^{k-1)} - \frac{2}{2k-1}
g^{(2} \eta^{k-2)a} h_\mu{}^a ]  \nonumber \\
\delta_1 \Phi_\mu{}^{(k)} &=& - \varepsilon^{ab(1} [ \eta^{k-1)a}
h_\mu{}^b + \xi^{k-1)a} \omega_\mu{}^b ] - (k+1) \alpha_{k+1}
h_\mu{}^a \xi^{a(k)} - \\
 && - \frac{k+1}{k-1} \alpha_k [ h_\mu{}^{(1} \xi^{k-1)} -
\frac{2}{2k-1} g^{(2} \xi^{k-2)a} h_\mu{}^a ] \nonumber
\end{eqnarray}
as can be easily seen comparing these expressions with the free ones
(\ref{arbd0}). But to find higher spin transformations for  graviton
requires more work. By straightforward but lengthy calculations we
obtain:
\begin{eqnarray}
\delta_1 \omega_\mu{}^a &=& \sum_{k=1}^{s-1} (-1)^{k+1} k
\varepsilon^{abc} [ \Omega_\mu{}^{b(k-1)} \eta^{c(k-1)} + \beta_k{}^2
\Phi_\mu{}^{b(k-1)} \xi^{c(k-1)} ] + \nonumber \\
 && + \sum_{k=2}^{s-1} (-1)^k k \alpha_k [ \frac{k+1}{k-1}
\Phi_\mu{}^{(k-1)} \eta^{a(k-1)} - \Phi_\mu{}^{a(k-1)} \eta^{(k-1)} +
\nonumber \\
 && \qquad \qquad \qquad + \Omega_\mu{}^{(k-1)} \xi^{a(k-1)} -
\frac{k+1}{k-1} \Omega_\mu{}^{a(k-1)} \xi^{(k-1)} ] - \\
 && - \gamma_0 (\eta_\mu \pi^a - \pi_\mu \eta^a)  \nonumber \\
\delta_1 h_\mu{}^a &=& \sum_{k=1}^{s-1} (-1)^{k+1} k \varepsilon^{abc}
[ \Phi_\mu{}^{b(k-1)} \eta^{c(k-1)} + \Omega_\mu{}^{b(k-1)}
\xi^{c(k-1)} ]  \nonumber
\end{eqnarray}
Thus we have constructed the cubic vertex and all corresponding
corrections to gauge transformations such that the resulting theory
is invariant in the linear approximation. Unlike the higher
dimensional case \cite{FV87}, \cite{FV87a}, the cubic vertex under
consideration does not require any higher derivative terms. Note also
that, as well as in the spin-3 case, for arbitrary $s$-case after
switching on gravitational interaction we still have the possibility
to take any of the massless, partially massless or flat limits.
In-particular, this allows one to consider interactions for such
partially massless fields that can possesses the specific properties
and have simpler structure in comparison with generic massive fields.
As we have already mentioned to go beyond linear approximation one has
to
introduce other fields into the system and till now what spectrum of
massive fields one has to consider to obtain consistent theory is an
open question that certainly deserves further study.

\section*{Conclusion}

In this paper we have studied gravitational interactions for massive
higher spin fields in three dimensions using frame-like gauge
invariant description.  At first, after providing a couple of
simplest interacting massless models illustrating our general
technics, we constructed gravitational interactions for massive
spin-3 field in linear approximation, e.g. found the cubic vertex
linear in gravitational field and quadratic in spin-3 field. Recall
that the linear approximation for any field do not depend on the
presence of any other fields in the system so that the very
existence of such cubic vertex is crucial for any further
investigations. Contrary to the massless case, and this may be the
main lesson from our work, we argue that the system gravity plus
massive spin-3 field is not closed. The main reason is that
Stueckelberg spin-2 and spin-1 fields are gauge fields themselves
having their own gauge symmetries so that the algebra must necessarily
be extended with corresponding generators. In this, even in the
massless limit we obtain a system containing graviton, massless spin-3
and massive spin-2 fields. In this, all this means that to
construct consistent theory we have to introduce some other field, the
most natural candidate being spin-4 one. But this almost certainly
will require introduction of even higher spins and so on. This
simplest example shows that the structure, spectrum of fields and
gauge algebra for massive and massless theories may be drastically
different. To proceed in this direction one first of all has to know
the structure of general cubic vertices for massive higher spins in
three dimensions that is still absent. Recall also, that for the
massive spin-3 case we have shown that after switching on
interactions these gauge invariant curvatures admit non-linear
deformations so that they transform covariantly under corrected gauge
transformations. This, in turn, suggests that the most natural route
to deal with such theories is some extension of Fradkin-Vasiliev
formalism adopted to three dimensions.

Then we managed to generalize these results to the case of massive
field with arbitrary integer spin. It is important that it turns out
to be possible without introduction of any higher derivative
corrections to the Lagrangian and/or gauge transformations and this is
one of the reason why three dimensional theories appear to be much
simpler that higher dimensional ones. Note also that even after
switching on gravitational interaction (at least in the linear
approximation) we still have the possibility to take any massless,
partially massless or flat limits.

\section*{Acknowledgments}
I.L.B and T.V.S are grateful to the grant for LRSS, project No
224.2012.2 and RFBR grant, project No 12-02-00121-a for partial
support. Work of I.L.B was also partially supported by RFBR-Ukraine
grant, project No 11-02-9045 and DFG grant, project LE 838/12-1.
T.V.S. acknowledge the partial support from RFBR grant No.
12-02-31713.  Work of Yu.M.Z was supported in parts by RFBR grant No.
11-02-00814.

\end{document}